\documentclass[twocolumn]{aastex62}

\usepackage{amsmath}
\usepackage{mathtools}
\usepackage{rotating}
\usepackage{booktabs}
\usepackage{upgreek}

\newcolumntype{L}[1]{>{\raggedright\arraybackslash}p{#1}}
\newcolumntype{C}[1]{>{\centering\arraybackslash}p{#1}}
\newcolumntype{R}[1]{>{\raggedleft\arraybackslash}p{#1}}

\submitjournal{ApJ}

\shorttitle{Experimental and theoretical studies of ${\rm D} + {\rm H}_3^+\rightarrow {\rm H}_2 {\rm D}^+ + {\rm H}$}
\shortauthors{Hillenbrand et al.}

\begin{document}

\title{Experimental and theoretical studies of the isotope exchange reaction ${\rm D} + {\rm H}_3^+\rightarrow {\rm H}_2 {\rm D}^+ + {\rm H}$}







\author[0000-0003-0166-2666]{P.-M.~Hillenbrand}
\altaffiliation{Present address: GSI Helmholtzzentrum f\"ur Schwerionen-\\forschung, D-64291 Darmstadt, Germany,\\ \url{p.m.hillenbrand@gsi.de}}
\affiliation{Columbia Astrophysics Laboratory, Columbia University, New York, NY 10027, U.S.A.}

\author[0000-0003-2483-8863]{K.~P.~Bowen} 
\affiliation{Columbia Astrophysics Laboratory, Columbia University, New York, NY 10027, U.S.A.}

\author[0000-0002-5557-7486]{J.~Li\'evin}
\affiliation{Service de Chimie Quantique et Photophysique, Universit\'e Libre de Bruxelles, B-1050 Brussels, Belgium}

\author[0000-0003-3326-8823]{X.~Urbain} 
\affiliation{Institute of Condensed Matter and Nanosciences, Universit\'e catholique de Louvain, B-1348 Louvain-la-Neuve, Belgium}

\author[0000-0002-1111-6610]{D.~W.~Savin} 
\altaffiliation{\url{savin@astro.columbia.edu}}
\affiliation{Columbia Astrophysics Laboratory, Columbia University, New York, NY 10027, U.S.A.}

\begin{abstract}
Deuterated molecules are important chemical tracers of prestellar and protostellar cores. Up to now, the titular reaction has been assumed to contribute to the generation of these deuterated molecules. We have measured the merged-beams rate coefficient for this reaction as function of the relative collision energy in the range of about 10~meV to 10~eV. By varying the internal temperature of the reacting ${\rm H}_3^+$ molecules, we found indications for the existence of a reaction barrier. We have performed detailed theoretical calculations for the zero-point-corrected energy profile of the reaction and determined a new value for the barrier height of $\approx 68$~meV. Furthermore, we have calculated the tunneling probability through the barrier. Our experimental and theoretical results show that the reaction is essentially closed at astrochemically relevant temperatures. We derive a thermal rate coefficient of $<1\times 10^{-12}$~cm$^3$~s$^{-1}$ for temperatures below $75$~K with tunneling effects included and below $155$~K without tunneling.

\end{abstract}

\keywords{astrochemistry - ISM: molecules - methods: laboratory - molecular data - molecular process}

\section{Introduction}
\label{sec:intro}

Deuterated molecules are important chemical tracers of the interstellar molecular clouds where stars form.  At the $\sim 10-20$~K typical of these environments, exoergic deuterium-substitution reactions go forward, but the endoergic hydrogen-substitution reverse reactions do not, due to the vibrational zero-point energy (ZPE) of a deuterated molecule lying below that of its H-bearing counterpart.  This fractionation process explains why, in cold environments, the observed abundance ratios of deuterated molecules relative to their H-bearing analogues are orders of magnitude larger than the galactic D/H ratio.  Since these findings became apparent, numerous astrochemical models have been developed to explain the observations \citep[an incomplete list of models includes:][]{Millar:1989:ApJ,Rodgers:1996:MNRAS,roberts_modelling_2000,walmsley_complete_2004,Flower:2006:AnA,aikawa_prestellar_2012,albertsson_new_2013,sipila_hd_2013,mcelroy_umist_2013,kong_deuterium_2015,lee_d/h_2015,majumdar_chemistry_2017}. 

A particularly important deuterated molecule for tracing the properties of the cold gas in star forming regions is H$_2$D$^+$.  Once the particle density of the cloud reaches $\sim 10^6$~cm$^{-3}$, heavy elements are predicted to freeze onto dust grains.  H$_3^+$ and its isotopologues are predicted to become the dominant carriers of positive charge, a role normally played by metals such as S$^+$ and Fe$^+$, along with C-, N-, and O-bearing molecules such as HCO$^+$, H$_3$O$^+$, and N$_2$H$^+$ \citep{vanderTak:2006:RSPTA}.  However H$_3^+$ and D$_3^+$ are not observable at such low temperatures as they have no dipole moment and lack a pure rotational spectrum.  Conversely, H$_2$D$^+$ and D$_2$H$^+$ have dipole moments and a pure rotational spectrum that can be excited at these temperatures.  For example, H$_2$D$^+$ has been observed in low-mass prestellar cores \citep{caselli_abundant_2003,vastel_detection_2004,vastel_distribution_2006,Pagani:2009:AnA,Friesen:2010:ApJ}, low-mass protostellar cores \citep{Stark:1999:ApJL,caselli_survey_2008,friesen_revealing_2014}, low-mass young stellar objects \citep{Stark:1999:ApJL,brunken_h2d+_2014}, and massive star-forming regions \citep{Harju:2006:AnA,Swift:2009:ApJ,pillai_h2d+_2012}.

In order to harness the full diagnostic power of H$_2$D$^+$ for cold and dense star-forming regions, accurate chemical abundance models are needed.  Measurements of the H$_2$D$^+$ abundance, combined with these models, can be used to determine the ionization fraction of the object.  This fraction sets the time-scale for the gas-phase chemistry of the gas, as ion-neutral reactions dominate such chemistry at these temperatures.  Additionally, reliable values for the electron number density relative to that of H$_2$, $x_e = n_e/n_{\rm H_2}$, are needed for calculating the electron-driven portion of the chemistry occurring in a cloud \citep{caselli_survey_2008}.  The quantity $x_e$ is approximately equal to the ionization fraction, assuming that the gas is neutral.  Lastly, the ionization and electron fractions determine the influence of magnetic fields on the dynamics of the object, especially for the ability of the ambient fields to support against gravitational collapse \citep{vanderTak:2006:RSPTA,Grenier:2015:ARAnA,kong_deuterium_2015}.

Of the six reactions identified as being key in the formation and destruction of H$_2$D$^+$ in cold and dense star-forming regions, two reactions involve HD, one involves D$_2$, and two involve atomic D \citep{albertsson_new_2013}.  Laboratory measurements exist for the reactions involving HD and D$_2$
\citep{adams_laboratory_1981,giles_study_1992,gerlich_h3+_2002,gerlich_deuterium_2002,hugo_h3+_2009}, and the rate coefficients are thought to be well understood.  The same cannot be said for the two reactions involving atomic D.  This is due to the experimental challenges of generating controlled and well quantified beams of atomic D \citep{Bruhns:2010:RScI}.   \citet{pagani_ortho-h2_2013} also highlighted that reactions with atomic D have a sizeable influence on the chemistry, especially at steady state when atomic D becomes important.  These studies suggest that our ability to reliably use H$_2$D$^+$ as a diagnostic for star-forming regions is hindered by the lack of accurate astrochemical data for the reactions of atomic D with H$_3^+$ forming H$_2$D$^+$ and with H$_2$D$^+$ destroying the molecule.

Here we focus on the H$_2$D$^+$ formation reaction

\begin{equation}
  \label{eq:D}
        {\rm D + H_3^+ \rightarrow H_2D^+ + H},
\end{equation}

\noindent which is exoergic by 51.51~meV \citep{ramanlal_deuterated_2004}.  The only published theoretical results for this reaction appears to be the classical dynamics study by \citet{moyano_interpolated_2004}, which do not include corrections for the isotope-dependent ZPE along the reaction path.  Their cross section results lie an order of magnitude below the Langevin value.  They hypothesize that this discrepancy might be reduced when quantum effects are taken into account.  However, Moyano et al.\ also predict that the reaction path possesses a barrier of $E_{\rm b} = 149$~meV and they do not account for the possible effects of tunneling.  So it is surprising that they report a nonzero cross section for collision energies below $E_{\rm b}$.

To help to resolve this issue, we have carried out laboratory measurements for Reaction~(\ref{eq:D}).  The measurements were performed using our dual-source, ion-neutral, merged-fast-beams apparatus \citep{oconnor_reaction_2015,de_ruette_merged-beams_2016}.  In addition, we have carried out new theoretical calculations for the ZPEs for all of the stationary points along the reaction path, giving an improved value for $E_{\rm b}$.  Using our combined experimental and theoretical results, we have developed a semi-empirical model for the reaction cross section, from which we have generated a thermal rate coefficient for Reaction~(\ref{eq:D}) for astrochemical models.

The rest of the paper is organized as follows: In Section~\ref{sec:experiment}, we briefly describe the experimental apparatus.  The measurement procedure and data analysis are highlighted in Section~\ref{sec:analysis}. Section~\ref{sec:theo} provides a theoretical description of the reaction path including the potential energy surface and the ZPE at all stationary points. The experimental results are presented in Section \ref{sec:mergedbeamsrate} and discussed in Section \ref{sec:discussion}. A summary is given in Section \ref{sec:summary}. Throughout the paper, uncertainties are quoted at a confidence level taken to be equivalent to a one-sigma statistical confidence level, unless otherwise noted.

\section{Experimental Apparatus}
\label{sec:experiment}

We have developed a dual-source, merged-fast-beams apparatus that enables us to study reactions between neutral atoms and molecular cations and to measure the charged daughter products.  The experimental apparatus and methodology have already been described in detail in \citet{oconnor_reaction_2015} and \citet{de_ruette_merged-beams_2016}.  We provide here only a brief description, emphasizing aspects that are new or specific to the present study.

\subsection{Neutral Beam}

The neutral beam is formed by photodetachment of a beam of D$^-$, the only bound level of which is the $^1S_0$ \citep{Rienstra-Kiracofe:2002:ChRv}.  The anions are generated using a Peabody Scientific duoplasmatron source, accelerated to form a beam with a kinetic energy $E_{{\rm D}^-}=12.00$~keV (5.96~keV~u$^{-1}$ or $1.07 \times 10^8$~cm~s$^{-1}$), and guided electrostatically into a Wien filter.  This charge-to-mass filter is used to select the desired D$^-$ beam and remove any other negatively charged particles extracted from the source.  Typical D$^-$ currents after the Wien filter were $3.7~\mu$A.  The D$^-$ beam is then directed into a photodetachment chamber by a series of electrostatic ion optical elements.

In this chamber, the anions pass through a floating cell at a voltage of $U_{\rm f}$.  Upon entering this cell, the anions assume an energy of $E_{\rm D^-} + eU_{\rm f}$, where $e$ is the elementary charge.  Within the floating cell, a few percent of the anions are photodetached by an $\sim 1$~kW laser beam at a wavelength of $\lambda=808$~nm (a photon energy of $h\nu = 1.53$~eV, where $h$ is Planck's constant and $\nu$ the photon frequency).  This energy lies close to the maximum of the photodetachement cross section \citep{mclaughlin_h_2017} and generates ground level atomic D via 

\begin{equation}
{\rm D}^-\left( ^1S_0 \right) + h\nu \rightarrow {\rm D}\left( ^2S_{1/2} \right) + e^-.
\end{equation}

\noindent We have previously used this technique to produce beams of neutral atomic H and D for studies of associative detachment \citep{Bruhns:2010:RScI,Kreckel:2010:Sci,bruhns_absolute_2010,Miller:2011:PhRvA,Miller:2012:PhRvA}.  Additional details can be found in \citet{oconnor_generation_2015}.

The energy of the neutral beam formed is $E_{\rm n}=E_{{\rm D}^-}+eU_{\rm f}$ and does not change upon leaving the floating cell.  The beam is collimated by a set of two 5-mm apertures separated by a distance of 3168~mm, one before and one after the photodetachment chamber.  The current before the first aperture was $3.3~\mu$A.  The remaining D$^-$ beam after the second aperture is electrostatically removed and directed into a beam dump, leaving a pure beam of ground level D that continues ballistically into the interaction region.

\subsection{Cation Beam}
\label{sec:ionbeam}

H$_3^+$ is generated using a Peabody Scientific duoplasmatron and extracted from the ion-source chamber through an aperture with a diameter of $d=0.25$~mm.  The cations are accelerated to form a beam of energy $E_{{\rm H}_3^+}=E_{\rm i} = 18.02$~keV or 5.96~keV~u$^{-1}$.  This energy has been selected to velocity match that of the neutral D beam for $U_{\rm f}=0~$~V.  The beam then passes through a Wien filter to select the desired H$_3^+$ and remove all other cations extracted from the source.  After the Wien filter, the H$_3^+$ beam is electrostatically directed into a set of two 5-mm collimating apertures separated by a distance of 3069~mm.  The current before the first aperture is typically $\approx 7~\mu$A.  The second aperture is followed by a $90^\circ$ electrostatic cylindrical deflector.  This deflector merges the cations onto the neutral beam (which passes through a hole in the outer plate of the deflector and then through the exit of the deflector into the interaction region).  Electrostatic ion optics after the last collimating aperture and before this merging deflector are used to maximize the overlap between both beams in the interaction region.

It is well known that duoplasmatrons form H$_3^+$ ions that are internally excited.  The lower limit for this excitation at $\sim 300$~K is due to the water-cooled walls of the duoplasmatron.  The upper limit is the predicted dissociation temperature of $\sim 4000$~K for H$_3^+$ in thermal equilibrium \citep{kylanpaa_first-principles_2011}.  Our previous studies of C and O reacting with H$_3^+$ inferred an internal temperature of $\sim 2500-3000$~K by comparing the measured thresholds for competing channels to those predicted theoretically \citep{oconnor_reaction_2015,de_ruette_merged-beams_2016}.

Here we adjusted the source operating conditions in order to vary the level of internal excitation.  The parameters that we varied were the pressure inside the duoplasmatron chamber, the arc current, the magnet current, and the filament current.  As we will discuss in more detail in Section~\ref{sec:mergedbeamsrate}, the level of H$_3^+$ internal excitation was most sensitive to the source pressure $p_{\rm s}$.

We estimated $p_{\rm s}$ from the pressure measured just outside the source chamber, $p_{\rm o}$.  The short distance between the source aperture and the turbomolecular pump on the system allows us to treat the problem as two chambers separated by an aperture.  Using the fact that $p_{\rm s}\gg p_{\rm o}$ and the basic formulae of molecular flow through an aperture gives \citep{ohanlon_book_2003}

\begin{equation}
\label{eq:ps_formula}
p_{\rm s}=p_{\rm o}\frac{4 S}{d^2} \sqrt{\frac{2 m_{{\rm H}_2}}{\pi k_{\rm B} T}}.
\end{equation}

\noindent Here $S=220$~l~s$^{-1}$, is the H$_2$ pumping speed of the turbomolecular pump, $m_{\rm H_2}$ is the H$_2$ mass, $k_{\rm B}$ is the Boltzmann constant, and $T=300$~K is the gas temperature.  Inserting these values into Equation~(\ref{eq:ps_formula}) yields

\begin{equation}
\label{eq:ps}
p_{\rm s} \approx 1 \times 10^4 p_{\rm o}.
\end{equation}

\noindent We operated the source at $p_{\rm o} = 0.72 - 7.2 \times 10^{-5}$~Torr, with the gauge calibrated for reading H$_2$.  This corresponds to $p_{\rm s}= 0.072-0.72$~Torr.

\subsection{Interaction Region}

The interaction region begins near the exit of the merger deflector, at $z=0$~mm, where $z$ is the distance along the overlap of the two beams.  The length of the interaction region is $L=1215$~mm and is set by the location of the entrance electrode of an electrostatic chicane, described below.  The profiles of the two parent beams are measured individually using rotating wire beam profile monitors \citep[BPMs;][]{seely_rotating_2008}, one near the beginning of the interaction region at $z=280$~mm and the other near the end at $z=1090$~mm.  The measured profiles are used to calculate the mean overlap factor of the parent beams, $\left< \Omega (z)\right>$, as well as the bulk angle between them, $\theta_{\rm bulk}$.  A Faraday cup can be inserted between the two BPMs to measure the cation beam current.  Typical ion currents at this point were $I_{\rm i} \approx 1.1~\mu$A.  For $U_{\rm f}=0$~V, daughter H$_2$D$^+$ ions form in the interaction region with an energy given by the initial $E_{\rm H_3^+} = 18.02$~keV plus $E_{\rm D} = 12.00$~keV and minus the energy of the replaced H atom, $E_{\rm H} = 6.01$~keV, resulting in  $E_{\rm H_2D^+} = 24.01$~keV.

\subsection{Signal Detection}

After the interaction region, the desired daughter products are separated from the parent beams by a series of electrostatic analyzers.  Using electrostatics allows us to analyze charged particles based on their kinetic energies.  The first analyzer is a chicane, consisting of a series of four pairs of parallel plate electrodes.  These deflect the charged particles in the horizontal direction.  For each pair of plates, $\ell$, one was set to a voltage $+U_\ell$ and the other to $-U_\ell$.  The orientation of the chicane deflection has been rotated $90^\circ$ from the configuration used in \citet{oconnor_reaction_2015} and \citet{de_ruette_merged-beams_2016}.

In our previous work we used the chicane to deflect the parent cation beam into a Faraday cup located after the first electrode, while guiding the product ions back onto the optical axis of the chicane, as defined by the neutral beam trajectory.  However, the geometry of the Faraday cup location requires a large mass difference between the parent and product ions.  This could not be fulfilled in the present experiment.  So the H$_3^+$ beam current, $I_{\rm i}$, was measured at the beginning and end of each setting of $U_{\rm f}$ during data acquisition, typically a 10~s interval.  The current was measured by applying a suitable voltage to the entrance electrode of the chicane.  At this voltage, the transmittance of the cation beam from the interaction region to the chicane Faraday cup was 100\%.  During the H$_2$D$^+$ signal-collection portion of the data-acquisition cycle, the voltage on the entrance electrode was set to transmit the product ions through the chicane.

The daughter H$_2$D$^+$ ions are directed by the chicane into the final analyzer.  This consists of a series of three 90$^\circ$ cylindrical deflectors, each with a bending radius of $\approx 137$~mm: a lower cylindrical deflector (LCD), a middle cylindrical deflector (MCD), and an upper cylindrical deflector (UCD).  The outer plate for each cylindrical deflector was set to a voltage of $+U_\ell$ and the inner plate to $-U_\ell$.  In contrast to our previous work, all three deflections are now arranged in one vertical plane.  The LCD and MCD together form a bend of $180^\circ$ and the UCD provides a 90$^\circ$ bend in the opposite direction.  A slit with a gap of 5~mm is positioned at the focus at the
exit of the MCD to help suppress background.  This background is due, in part, to H$_3^+$ ions that make their way out of the chicane and into the final analyzer.  The rear deflector of the chicane is used to correct for slight misalignments of the beam perpendicular to the vertical deflection plane of the final analyzer.  The transmittance from the interaction region to the exit of the UCD was measured at $T_{\rm a}=90\pm5\%$ using a proxy cation beam at the energy of the signal ions and a Faraday cup after the exit of the UCD.

\begin{figure*}[t]
\gridline{\fig{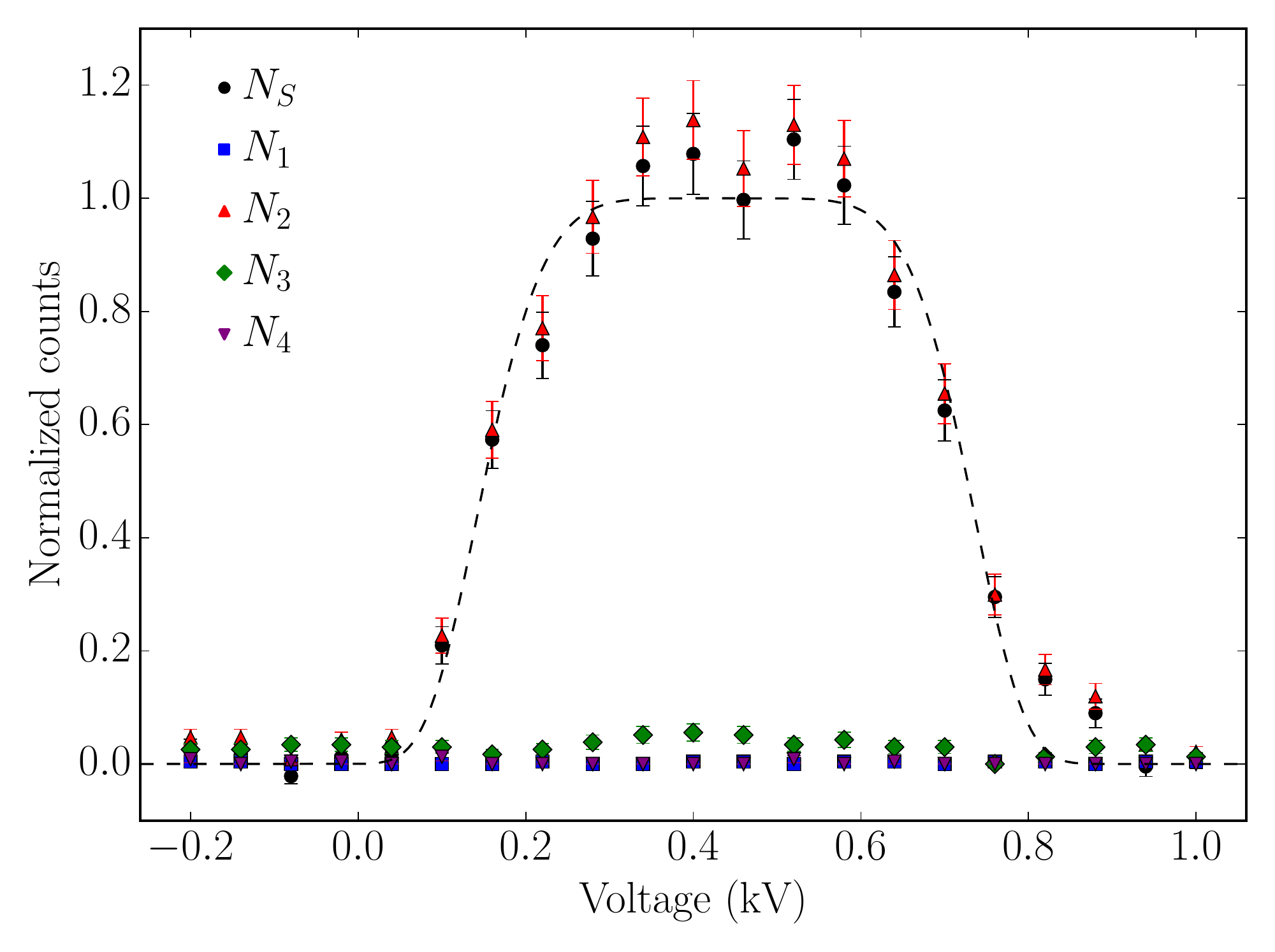}{0.47\textwidth}{(a) Rear chicane}\fig{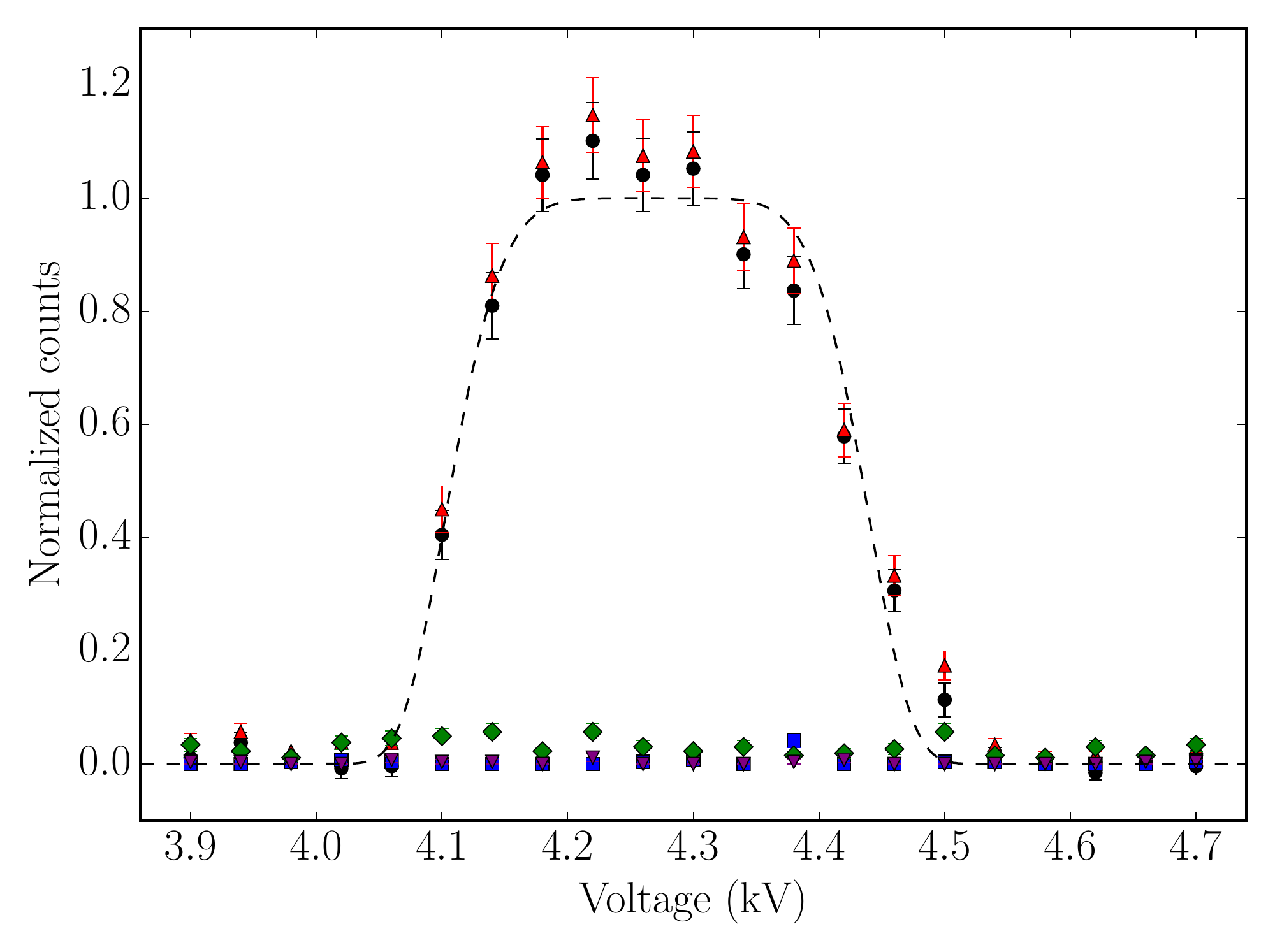}{0.47\textwidth}{(b) LCD}}
\gridline{\fig{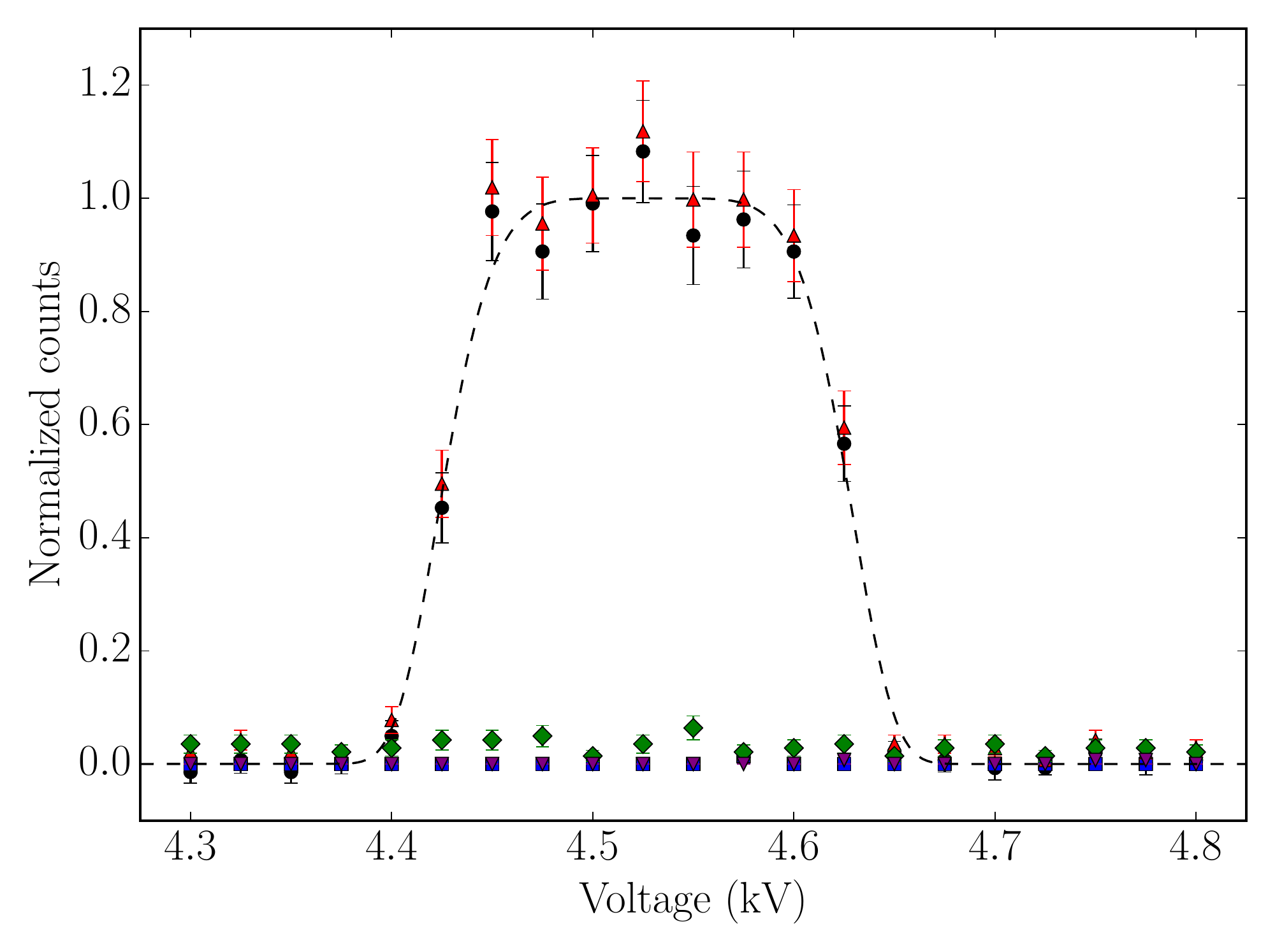}{0.47\textwidth}{(c) MCD}\fig{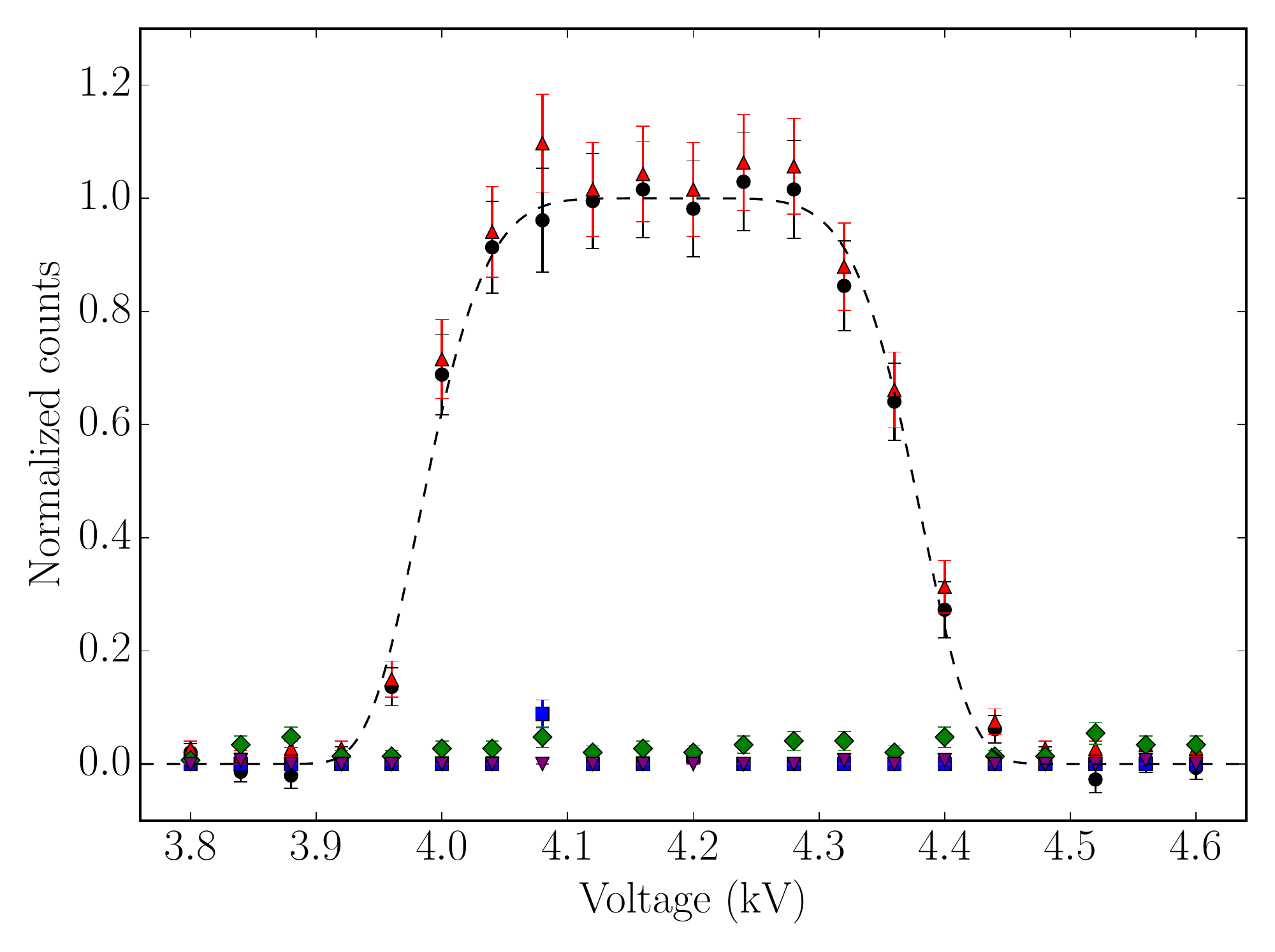}{0.47\textwidth}{(d) UCD}}
\caption{Voltage scans of the electrostatic analyzers for $U_{\rm
    f}=0$~V for the (a) rear chicane, (b) LCD, (c) MCD, and (d) UCD.
  For these scans, each voltage was set to $|U_\ell|=0.439$, 4.271, 4.526, and
  4.182~kV, respectively, when not being scanned.  Shown are the
  normalized counts for the different phases of the measurement cycle,
  which provide unambiguous background subtraction: For $N_1$ (blue
  squares), only the D beam is on.  For $N_2$ (red upward triangles),
  both beams are on.  For $N_3$ (green diamonds), only the H$_3^+$
  beam is on.  For $N_4$ (purple downward triangles), both beams are
  off.  The background corrected signal (black circles) is given by
  $N_S$ (see Section~\ref{subsec:Signal}).  The dashed lines are
  normalized fits using a modified Gaussian function
  $\exp[-(U-U_0)^6/(2\sigma^6)]$, where $U$ is the applied voltage,
  $U_0$ the central voltage, and $\sigma$ a fitting parameter.  These
  fits are given as a guide to the eye.}
\label{fig:scans}
\end{figure*}

Product ions are counted after the exit of the UCD using a channel electron multiplier (CEM) with an efficiency of $\eta=99\pm3\%$.  A repeller grid is located in front of the CEM and biased negatively to repel electrons.  The geometric transmittance of this grid is $T_{\rm g}=90\pm1\%$.  Typical H$_2$D$^+$ signal count rates were $S \approx 20$~s$^{-1}$.  The voltages on the chicane exit electrode, LCD, MCD, and UCD were scanned to determine the optimal settings for signal collection.  Representative scans are shown in Figure~\ref{fig:scans} for $U_{\rm f}=0$~V.

The trajectory of the H$_2$D$^+$ products is determined by the voltages applied to the four deflectors of the chicane and the three deflectors of the final analyzer.  During data acquisition, we typically scan $U_{\rm f}$ to vary $E_{\rm r}$.  This also varies $E_{\rm H_2D^+}$ and the various deflector voltages must be scaled accordingly.  Denoting the deflectors from the entrance electrode of the chicane to the UCD as $U_\ell$, with $\ell=1-7$, we scale $U_\ell$ versus $U_{\rm f}$ as

\begin{equation}
  \label{eq:scaling}
  U_\ell\left( U_{\rm f}\right)
  =
  U_\ell(0) \left(1+ \frac{eU_{\rm f}}{E_0} \right).
\end{equation}

\noindent Here $E_0=24.01$~keV is the H$_2$D$^+$ energy for $U_{\rm f}=0$~V.  These settings were routinely confirmed with signal scans similar to those shown in Figure~\ref{fig:scans}.

\subsection{Neutral Current}
\label{sec:neutral}

The neutral D beam travels ballistically from the interaction region, through the chicane, into the entrance aperture of the LCD, through an exit hole in the outer plate of the LCD, and into the neutral detector.  The transmission of the neutral beam from the interaction region to the neutral detector, dubbed the neutral cup (NC), is $T_{\rm n} = 95 \pm 3\%$, as measured using proxy ion beams.

The neutral particle current, $I_{\rm n}$, is measured in amperes.  The neutral beam strikes a target inside the neutral cup, which is configured to collect the resulting secondary emission of negative particles from the target \citep{bruhns_absolute_2010}.  This neutral cup can also be configured externally to serve as a Faraday cup for ion current measurements.  The measured neutral current is given by

\begin{equation}
  \label{eq:gamma}
I_{\rm n}=\frac{I_{\rm NC}}{\gamma T_{\rm n}},
\end{equation}

\noindent where $I_{\rm NC}$ is the negative particle current measured by the neutral cup and $\gamma$ is the mean number of negative particles emitted by a neutral particle striking the target.  For our work here, typical neutral D currents were $I_{\rm n}=43$~nA.

We measured the value of $\gamma$ using collisional detachment of the D$^-$ beam on a gas target, in this case the interaction region filled with Ar at a typical pressure of $6 \times 10^{-4}$~Torr, using a pressure gauge calibrated for Ar.  The resulting D beam was measured in the neutral cup.  The remaining D$^-$ beam was deflected by the LCD into the MCD (which had no applied voltage), passed through a hole in the outer plate of the MCD, and was measured in a Faraday cup, dubbed the upper cup (UC).  The transmittance of the D$^-$ beam from the interaction region to the upper cup was measured to be $T_{\rm u} = 65\pm2\%$.  Baseline measurements were also carried out for a residual gas pressure of $8 \times 10^{-8}$~Torr, using the same Ar-calibrated gauge.

The resulting value of $\gamma$ is given by

\begin{equation}
\gamma = 
\left( 1+ \frac{\sigma_{\rm DED}}{\sigma_{\rm SED}} \right) 
\frac{\Delta I_{\rm NC}}{\Delta I_{\rm UC}} 
\frac{T_{\rm u}}{T_{\rm n}}.
\end{equation}

\noindent $\Delta I_{\rm NC}$ and $\Delta I_{\rm UC}$ represent the measured current changes in the neutral cup and upper cup, respectively.  Each of these needs to be corrected for by the transmittance from the interaction region to the corresponding cup: $T_{\rm n}$ and $T_{\rm u}$, respectively.  We also accounted for the unmeasured D$^+$ cations generated by double electron detachment (DED) of D$^-$ on Ar.  The ratio of the DED cross section, $\sigma_{\rm DED}$, compared that for single electron detachment (SED), $\sigma_{\rm SED}$, is $\sigma_{\rm DED}/\sigma_{\rm SED}=3.5\%$.  This is based on the compilation of \citet{phelps_collisions_1992} and assumes that the cross sections for D$^-$ on Ar are the same as those for H$^-$ at matched velocities.

We measured $\gamma$ over several measurement series spread out over a number of weeks and also for a range of values for $E_{\rm n}$.  At 12.00~keV, we found $\gamma=1.6\pm0.1$.  As a function of $E_{\rm n}$, the data showed a small linear dependence of 

\begin{equation}
\gamma(E_{\rm n} \left[ {\rm keV} \right])=0.113 E_{\rm n} + 0.244,
\end{equation}

\noindent within the energy range of $E_{\rm n}=11.1-13.0$~keV studied here.  We accounted for this variation in the data analysis of our results.

\section{Measurement and Analysis}
\label{sec:analysis}

We begin by explaining the signal determination (Section~\ref{subsec:Signal}), followed by the data acquisition procedure (Section~\ref{subsec:measurement}), which has been enhanced since the work of \citet{oconnor_reaction_2015} and \citet{de_ruette_merged-beams_2016}.  Next we discuss the relative translational energy scale of the collision (Section~\ref{subsec:energy}).  Then we review how we evaluate the corresponding merged-beams rate coefficient (Section~\ref{subsec:rateeval}).  Again, we provide here only a brief description, emphasizing aspects that are new or specific to the present study.  Additional details can be found in \citet{bruhns_absolute_2010} and \citet{oconnor_reaction_2015}.

\subsection{Signal Determination}
\label{subsec:Signal}

In order to extract the desired signal, the two beams are chopped on and off, but out of phase with one another.  This enables us to unambiguously subtract the various backgrounds.  The chopping cycle is governed by the laser operating in a square-wave mode: on for 5~ms and then off for 5~ms.  The H$_3^+$ beam is electrostatically chopped with the same time structure, but delayed by a phase shift of 2.5~ms.  This chopping cycle is repeated for 10~s at a given value of $U_{\rm f}$.

In the first phase of this chopping cycle, only the D beam is on and the counts are denoted by $N_1$.  In the second phase, both beams are on and the counts are $N_2$.  In the third phase, only the H$_3^+$ beam is on and the counts are $N_3$.  In the last phase, both beams are off and the counts are $N_4$.  The desired signal counts $N_s$ are given by

\begin{equation}
  N_S = N_2 - N_1 - N_3 + N_4
\end{equation}

\noindent and the corresponding statistical uncertainty by

\begin{equation}
  \delta N_S = \left(N_1 + N_2 + N_3 + N_4\right)^{1/2}.
\end{equation}

\noindent The signal rate $S$ is given by dividing $N_S$ by the corresponding integration time of $\tau = 2.5$~s at each step in the chopping pattern.  The fractional statistical uncertainty in $S$ is given by $\delta N_S/N_S$.

\subsection{Data Acquisition Procedure}
\label{subsec:measurement}

Each data run typically consists of 10 scans of $U_{\rm f}$ ($i=1-10$), which is swept through a series of 20 voltage steps ($j=1-20$) for each scan.  A run corresponds to about one hour and is comparable to the timescale over which both beams are stable.

The $U_{\rm f}$ scan ranges used here were $-900$~V to 1000~V, $-450$~V to 500~V, and $-225$~V to 250~V.  Measurements of the ion and neutral beam profiles are performed independently at the beginning and end of each sweep.  For the neutral beam measurements, we found no significant variation over the range scanned in $U_{\rm f}$.  So we set $U_{\rm f}=0$~V for the neutral beam profile measurements.  The data presented below represent the average of various accumulated data runs over the three $U_{\rm f}$ ranges listed above.

Signal is collected within a predefined sweep range for $U_{\rm f}$ by automatically incrementing the floating cell voltage every 10~s.  The voltages of the chicane and the final analyzer are scaled synchronously with each step of the floating cell voltage, as given by Equation~(\ref{eq:scaling}).  This configuration is to be contrasted with our earlier work where data were collected at just one floating cell voltage for each data run \citep{oconnor_reaction_2015,de_ruette_merged-beams_2016}.

As mentioned earlier, it is not possible to set the voltages on the chicane to simultaneously direct the H$_3^+$ into the chicane Faraday cup and transmit the product H$_2$D$^+$ into the final analyzer.  To overcome this, we measure the H$_3^+$ current, $I_{\rm i}$, before and after each 10-s increment at a given $U_{\rm f}$ using the chicane Faraday cup as described earlier.  We have confirmed that the ion beam is sufficiently stable over a 10~s increment to
justify this.  

\subsection{Relative Translational Energy and Beam Overlap}
\label{subsec:energy}

The relative translational energy $E_{\rm r}$ in the center-of-mass system for mono-energetic beams intersecting at an angle $\theta$ is given by
\citep{brouillard_measurement_1983}

\begin{equation}
\label{eq:energy}
E_{\rm r} = 
\mu 
\left(
\frac{E_{\rm n}}{m_{\rm n}} + \frac{E_{\rm i}} {m_{\rm i}}
- 2\sqrt{\frac{E_{\rm n}E_{\rm i}}{m_{\rm n}m_{\rm i}}}\cos\theta
\right).
\end{equation}

\noindent Here $m_{\rm n}=2.015$~u and $m_{\rm i}=3.023$~u are the masses of the D atom and the H$_3^+$ ion, respectively \citep{linstrom_nist_2018}.  The reduced mass is defined as

\begin{equation}
\label{eq:mass}
\mu = \frac{m_{\rm n} m_{\rm i}}{m_{\rm n} + m_{\rm i}}.
\end{equation}

\noindent For our work here, we have $\mu = 1.209$~u.  The corresponding relative velocity is

\begin{equation}
  \label{eq:vr}
  v_{\rm r}=\sqrt{\frac{2E_{\rm r}}{\mu}}.
\end{equation}

In our experiment the two beams interact over a range of angles and with a spread in kinetic energies.  The former is determined by $\theta_{\rm bulk}$ between the two beams combined with the divergences of each beam.  The latter is determined by the $\pm 10$~eV energy spread of each source.  We have calculated the resulting $E_{\rm r}$ using a Monte Carlo particle ray tracing as described in \citet{bruhns_absolute_2010} and \citet{oconnor_reaction_2015}.  These simulations were adjusted to match the constraints from the various collimating aperture dimensions and locations in the apparatus as well as from the measured beam profiles.  In specific, the simulations were adjusted to reproduce the measured typical bulk angle of $\theta_{\rm bulk}=0.39\pm0.19$~mrad, beam profiles, overlaps, and the overlap integral of $\left<\Omega(z)\right>=2.81 \pm 0.19~\rm{cm}^{-2}$, which was calculated from the beam profiles measured along the interaction region as described by \citet{bruhns_absolute_2010} and \citet{oconnor_reaction_2015}.

The simulations also yield a histogram of relative translational energies throughout the interaction volume.  We take the mean of this distribution as our experimental $E_{\rm r}$ and the one-sigma spread of the histogram, $\Delta E_{\rm r}$, as our relative energy uncertainty.  The resulting distribution of $E_{\rm r}$ for a given $U_{\rm f}$ is nearly Maxwellian for low values of $\left|U_{\rm f}\right|$ and converges to a Gaussian distribution for larger values of $\left|U_{\rm f}\right|$.

Additional fine-tuning of the $E_{\rm r}$ scale is achieved by comparing the results measured when the neutrals are faster than the ions ($U_{\rm f} > 0$~V) to when they are slower ($U_{\rm f} < 0$~V).  The results should be symmetric in magnitude around $U_{\rm f}=0$~V.  We find that the expected symmetry requires applying a small correction of $+6$~V to $U_{\rm f}$.  We attribute this to slight differences in the plasma potentials between the D$^-$ and H$_3^+$ duoplasmatron sources.  Taking this into account in our simulations results in a calculated minimum experimental $E_{\rm r} = 9 \pm 7$~meV, corresponding to a translational temperature of $\approx 70$~K.  The highest collisions energies studied correspond to $10.8\pm0.1$~eV and $11.8 \pm 0.1$~eV for $U_{\rm f}=-0.9$~kV and $1.0$~kV, respectively.

\subsection{Merged-Beams Rate Coefficient}
\label{subsec:rateeval}

We measure the cross section, $\sigma$, for Reaction~(\ref{eq:D}) times the relative velocity, $v_{\rm r}$, between the collidors convolved with the energy spread of the experiment.  The merged-beams rate coefficient and corresponding uncertainty for a given $U_{\rm f}$ scan $i$ and voltage step $j$ is given by

\begin{equation}\label{eq:rateeval}
\langle \sigma v_{\mathrm r} \rangle_{i,j}
= 
\frac{N_{S_{i,j}} \pm \delta N_{S_{i,j}}}{\tau}
\frac{1}{T_{\mathrm a}T_{\mathrm g}\eta} 
\frac{e^2v_{\mathrm n}v_{\mathrm i}}{I_{\mathrm n}I_{\mathrm i}} 
\frac{1}{L\langle \Omega (z) \rangle}.
\end{equation}

\noindent Here, the velocities $v_{\mathrm n}$ and $v_{\mathrm i}$ are of the neutral and molecular ion beams, respectively, and are calculated using the corresponding beam energies.  The other variables have been defined previously.  We measure each of the quantities on the right-hand side of Equation~(\ref{eq:rateeval}), enabling us to generate absolute results, independent of any normalization.

Typical values of the experimental parameters going into Equation~(\ref{eq:rateeval}) and their uncertainties are summarized in Table \ref{tab:errors}.  The neutral current is given by the average over the 10-s period $j$ and the ion current by the average of the measurements before and after this period.  $\langle \Omega(z) \rangle$ is taken from the average of all overlap measurements in a given data run, typically eleven.  Those quantities that varied between the steps of a scan are grouped under ``{\it Non-constants''} in Table~\ref{tab:errors} and those that remained constant throughout all runs are grouped under ``{\it Constants}''.  

\begin{table}[t]	
\begin{center}
\caption{Typical experimental values for Equation~(\ref{eq:rateeval}) with corresponding uncertainties. The total systematic uncertainty (excluding the statistical error) is calculated by treating each individual uncertainty as a random sign error and adding all in quadrature.}
\label{tab:errors}
\tabcolsep=0.1cm
\begin{tabular}{lccc}
\hline
\hline
Source                   & Symbol 	          & Value		& Uncertainty \\
& & & (\%) \\
\hline 
{\it Non-constants:} & & & \\
Signal rate    	& $S$                     & 20 s$^{-1}$               & $\leq$ 9\\
\ \ (statistical) & & & \\
D velocity       		& $v_{\mathrm n}$           & $1.07 \times 10^8$~cm~s$^{-1}$ 	& $\ll$1   \\
D current			& $I_{\mathrm n}$           & 43 nA               & 5        \\
H$_3^+$ current	 		& $I_{\mathrm i}$	          & 1.1 $\mu$A          & 5        \\
Overlap factor  		& $\left<\Omega(z)\right>$  & 2.8 cm$^{-2}$      & 10       \\
Neutral detector	        & $\gamma$ 		 & 1.6      		& 6 \\
\ \ efficiency & & & \\
\\
{\it Constants:} & & & \\
H$_3^+$ velocity 		& $v_{\mathrm i}$	  	& $1.07 \times 10^8$~cm~s$^{-1}$ 	& $\ll$1   \\
Analyzer	        & $T_{\mathrm a}$ 	& 0.90     		& 5 \\
\ \ transmission & & & \\
Grid		& $T_{\mathrm g}$ 	& 0.90     		& 1 \\
\ \ transmission & & & \\
Neutral	        & $T_{\mathrm n}$ 	& 0.95    		& 3 \\
\ \ transmission & & & \\
CEM efficiency			& $\eta$            	& 0.99     		& 3 \\
Interaction		& $L$               	& 121.5 cm 		& 2 \\
\ \ length & & & \\
\hline 
\multicolumn{3}{l}{Total systematic uncertainty} & 15\\
\multicolumn{3}{l}{\ \ (excluding the signal rate)} & \\
\hline
\end{tabular}
\end{center}
\end{table}

In order to calculate $\langle \sigma v_{\rm r} \rangle_j$ and the corresponding uncertainty for a given data run, we used the unweighted average of the results from all voltage scans $i$, given by

\begin{equation}
  \langle \sigma v_{\rm r} \rangle_j
  =
  \frac{\Sigma_{i=1}^{i_{\rm max}} \langle \sigma v_{\rm r} \rangle_{i,j}}{i_{\rm max}}.
\end{equation}

\noindent Various data runs were combined using a statistically-weighted average of all measured $\langle \sigma v_{\rm r} \rangle_j$ at the same $U_{\rm f}$ step
\citep[e.g.,][]{oconnor_reaction_2015}.  Finally, the values of $E_{\rm r}$ and $\Delta E_{\rm r}$ were assigned to each value of $U_{\rm f}$, based on the average overlap and bulk angle from all data runs, as described in Section \ref{subsec:energy}.

\section{Theoretical considerations}
\label{sec:theo}

\subsection{Energy Profile of the Reaction Path}

The 6-dimensional Born-Oppenheimer (BO) electronic potential energy surface (PES) of the H$_4^+$ cation defines the energy landscape  governing the dynamics of the isotopic exchange reaction

\begin{equation}
  \label{eq:theo}
{\rm X} + {\rm H}_3^+ \rightarrow {\rm XH}_2^+ + {\rm H},
\end{equation}

\noindent where X~=~H or D.

For the X/H exchange reaction, a wave packet propagating on this PES will follow a route connecting the entrance channel to the exit channel.  Along this path the system will cross stationary points located on this surface (global/local minima and transition states).  The relative energies of these critical points and the minimum energy path linking them define the BO-energy profile of the reaction.  This profile is useful for discussing our experimental results and how they are affected by the presence of the potential energy barrier along the reaction path.

The height of the barrier with respect to the entrance channel corresponds to the minimum energy classically required to observe reactive trajectories. However, quantum mechanics requires that the total internal energy of the system be greater than or equal to its vibrational ZPE.  It is therefore necessary to add the ZPE values of the different stationary points to the corresponding BO energies, leading to a vibrationally adiabatic minimum energy path \citep{Jankunas2014}.  We refer to this below as the ZPE-corrected energy profile.  The corrected barrier height can then be used to predict the minimum collision energy at which a nonzero cross section would be observed, in absence of quantum tunneling.  Note that the BO PES and the resulting energy profile are independent of the nuclear masses and are identical for the X~=~H or D reactions.  However, the corresponding ZPE-corrected profiles acquire the mass-dependency of the vibrational energies and are thus not identical for X~=~H and D.

Below, we determine the ZPE-corrected energy profiles from \textit{ab initio} calculations.  For this we build on previously published \textit{ab initio} work characterizing the H$_4^+$ BO PES \citep{Jiang1998,Alvares-Collado1995,moyano_interpolated_2004,Alijah2008,Sanz-Sanz2013}.

\subsection{Topography of the H$_4^+$ PES}

The global topography of the H$_4^+$ PES is well known.  The stationary points have been characterized (energies and geometries) at different levels of \textit{ab initio} theory and the convergence toward exact energies has been carefully investigated \citep{moyano_interpolated_2004, Alijah2008, Sanz-Sanz2013}. Harmonic vibrational frequencies and the corresponding ZPE values have been calculated, but only for the H$_4^+$  isotopologue \citep{Alijah2008, Sanz-Sanz2013}. Global analytical PESs have also been interpolated from \textit{ab initio} points,  first by \citet{moyano_interpolated_2004} at a medium level of theory and later by \citet{Sanz-Sanz2013} at a higher level.  These calculations predict the energy path describing the X/H exchange reaction.  

\begin{figure}[t!]
	\includegraphics[scale=0.5]{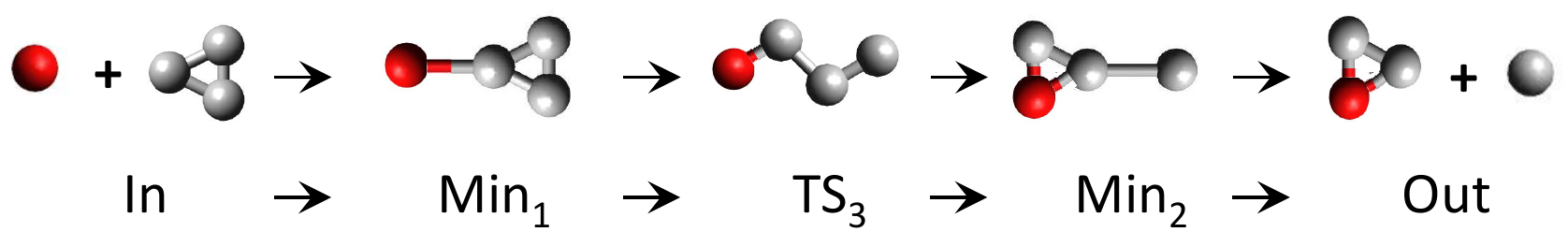}
	\caption{\label{fig:theo1} Molecular rearrangements along the minimum energy reaction path of the exchange reaction involving an X atom (H or D, shown in red) colliding with H$_3^+$. The molecular plane is maintained along the entire path.  See text for details.}
\end{figure}

The successive molecular rearrangements occurring along the reaction coordinate during the exchange reaction are illustrated schematically in Figure~\ref{fig:theo1}.  The X atom (H or D) collides with H$_3^+$ and forms an equilibrium structure, Min$_1$, in which X is weakly bound to the H$_3^+$ moiety. The X/H exchange is made possible by the transformation of Min$_1$ into a BO-equivalent minimum structure, Min$_2$, where X is now embedded into a triangular XH$_2^+$ structure, to which an H atom is weakly bound. This transformation implies the passage through a transition state, TS$_3$, using the nomenclature of the previous works cited above.  Dissociation of Min$_2$ leads to the final products of XH$_2^+$ and H.  Note that the permutational symmetry arising from the identical hydrogen atoms of H$_3^+$ allows the wave packet to explore with an equal probability several pathways equivalent to the one depicted in Figure~\ref{fig:theo1}.  As the permutational properties do not depend on X, it follows that both the H/H and D/H exchange reactions share the same pathways.  However, as we show below, the ZPE-corrected energies are affected by the D/H isotopic substitution.

\subsection{\textit{Ab Initio} Results}
\label{sec:theoresults}

\begin{figure}[t!]
	\begin{center}
	\includegraphics[scale=0.6]{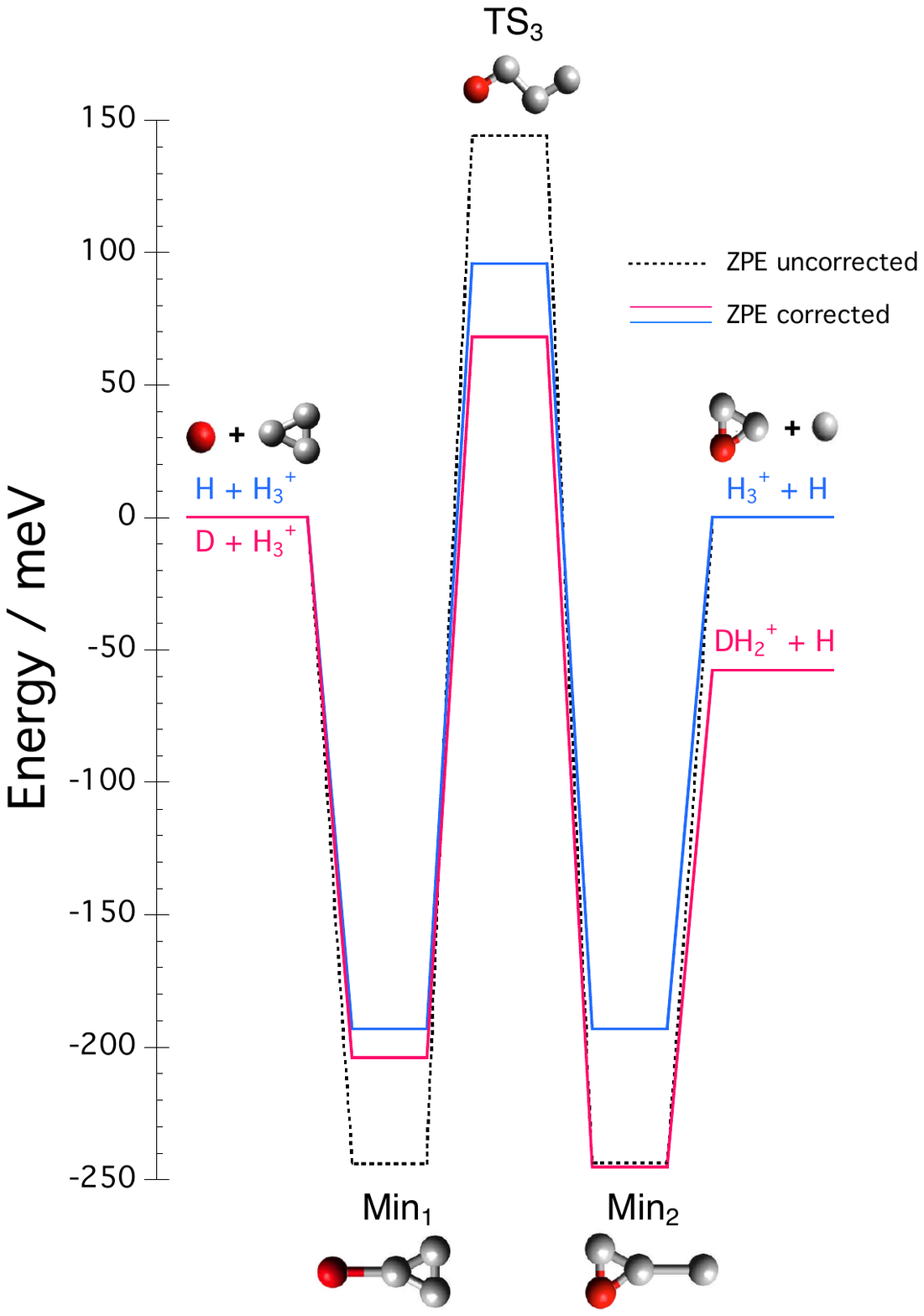}
	\caption{\label{fig:theo2} Our calculated energy profiles of the exchange reactions.  The black dashed line is the BO energy profile (ZPE uncorrected) and the colored lines correspond to the ZPE-corrected profiles (blue solid line for X~=~H and red solid line for X~=~D).  The molecular structures are shown at all stationary points, with the X atom colored in red.  See text for details.}
	\end{center}
\end{figure}

The ZPE-corrected energy profiles for the X~=~H and D reactions have been calculated using \textit{ab initio} theory carried out with the Molpro program package \citep{Werner2012,MOLPRO}.  The method we have used consists of a complete active space self consistent field (CASSCF) calculation \citep{Werner1985, Knowles1985} followed by an internally contracted multi-reference configuration interaction (ic-MRCI) calculation \citep{Werner1988,Knowles1988}.  This highly correlated CASSCF/ic-MRCI approach was also adopted in previous works by \citet{Alijah2008} and \citet{Sanz-Sanz2013}.  For our work, we used a large active space including 16 molecular orbitals and the extended Dunning's augmented correlation-consistent polarized quintuple-zeta (aug-cc-pV5Z) basis set \citep{Dunning1989, Kendall1992}, producing energies close to the corresponding full configuration interaction (FCI) limit (within $2\times10^{-7}$ $E_{\rm h}$, where $E_{\rm h}$ is the Hartree energy). The equilibrium geometries and the  harmonic vibrational frequencies of the different stationary points were calculated for the X~=~H and D isotopologues.

Our results for the calculated energy profiles, one without the harmonic ZPE corrections and two with the corrections, are shown in Figure~\ref{fig:theo2}.  The BO electronic energy for the entrance channel is the same for both X~=~H and D.  The corresponding $-1.8436004 E_{\rm h}$ has been subtracted out and the difference expressed in meV, so as to better show the relative energy variations in the stationary points and the exit channel.

\begin{table}[t!]
	\begin{center}
		\caption{Characteristic energies for the profiles in Figure~\ref{fig:theo2}. The potential well, barrier height, and exoergicity (all given in meV) correspond respectively to the energies of the minima, transition state, and exit channel, with respect to the entrance channel. The resulting barrier height, $E_{\rm b}$, and the relative energy of the exit channel, $\Delta E_{\rm zp}$, are in bold.} 
		\label{tab:theo1}
		\tabcolsep=0.1cm
		\begin{tabular}{llccc}
			\hline \hline
			Property		         & 		   & ZPE-uncorrected\tablenotemark{a} & \multicolumn2c{ZPE-corrected\tablenotemark{a}} \\
			\cline{4-5}
			                                 &  		   &                                  &  X=H    & X=D \\
			\hline
			\decimals
			Potential well                   & Min$_1$     & -243.98                          & -193.19 & -204.66 \\
			                                 & Min$_2$     & -243.98     			          & -193.19 & -245.74 \\
			                                 & Min$_{1,2}$  & -239.41\tablenotemark{b}         & -       & -	\\
			                                 & Min$_{1,2}$  & -243.95\tablenotemark{c}         & -       & -	\\						
			\hline
			Barrier height                   & TS$_3$      &  143.99	                  & 95.81   & \textbf{67.98} \\
			                                 & TS$_3$      &  149.25\tablenotemark{b}	  & -       & -	\\
			                                 & TS$_3$      &  144.00\tablenotemark{c}         & -       & -	\\
			\hline
			Exoergicity                      &             &  0.00                            & 0.00    & \textbf{-57.80} \\
		                                         &             &                                  &         & -54.10\tablenotemark{d} \\
		                                         &             &                                  &         & -51.51\tablenotemark{e} \\
			\hline \hline	
		\end{tabular}
	\end{center}
	\tablenotetext{a}{From this work unless otherwise specified.}
	\tablenotetext{b}{From \citet{moyano_interpolated_2004}.}
	\tablenotetext{c}{Calculated from \citet{Sanz-Sanz2013}.}
	\tablenotetext{d}{Includes the difference in electron binding energy for atomic H and D of 3.70~meV \citep{kramida_nist_2018}, as was done by \citet{ramanlal_deuterated_2004}.}
	\tablenotetext{e}{From \citet{ramanlal_deuterated_2004}.}
\end{table}

\begin{table*}[t!]
	\begin{center}
		\caption{Harmonic vibrational frequencies and corresponding ZPE values from \textit{ab initio} calculations. The imaginary frequency of the transition state, $\upomega_{\rm im}$, for Reaction~(\ref{eq:D}) is marked in bold.} 
		\label{tab:theo2}
		\tabcolsep=0.1cm
		\begin{tabular}{lccccccccc}
			\hline \hline
			Structure &  Property & Reference\tablenotemark{a} & ZPE (cm$^{-1}$) & \multicolumn6c{Harmonic frequencies (cm$^{-1}$)}  \\
			\cline{5-10}
            & & & $E_{\rm zp}$ & $\upomega_1$ & $\upomega_2$ & $\upomega_3$ & $\upomega_4$ & $\upomega_5$ & $\upomega_6$ \\
			\hline 
			H$_4^+$ & Min$_{1,2}$ &  & 4965.2 & 597.0 & 607.2 & 783.3 & 2221.6 & 2278.4 & 3442.8 \\
			& Min$_{1,2}$ & \citet{Alijah2008} & 4969 & 596 & 608 & 786 & 2224 & 2280 & 3443 \\
 			& Min$_{1,2}$ & \citet{Sanz-Sanz2013}\tablenotemark{b} & 4955 & 597 & 607 & 764\tablenotemark{c} & 2221 & 2278 & 3443 \\
 			& TS$_3$ &  & 4167.0 & 942.4i & 504.4 & 976.3 & 2009.0 & 2080.0 & 2764.3 \\
 			& TS$_3$ & \citet{Alijah2008} & 4168 & 950i & 506 & 974 & 2011 & 2079 & 2765 \\
 			& TS$_3$ & \citet{Sanz-Sanz2013}\tablenotemark{b} & 4167 & 942i & 505 & 977 & 2009 & 2080 & 2764 \\
			\hline
			H$_3$D$^+$ & Min$_1$ &  & 		4872.7 & 476.9 & 581.5 & 762.4 & 2203.9 & 2277.9 & 3442.8 \\
 			& Min$_2$ &  & 4541.8 & 555.3 & 597.7 & 769.1 & 1961.4 & 2170.7 & 3029.5 \\
 			& TS$_3$ &  & 3942.5 & \textbf{875.9i} & 472.1 & 875.4 & 1908.0 & 2010.0 & 2619.5 \\
			\hline
			H$_3^+$ & & & 4491.5 & 2773.2 & 2773.2 & 3436.5 & & & \\
 			&  & \citet{lie_vibrational_1992} & 4494.3 & 2774.9 & 2774.9 & 3438.8 &  &  &  \\
			&  & & 4555.6\tablenotemark{d} &  &  &  &  &  &  \\
			\hline
			H$_2$D$^+$ & & & 4089.4 & 2409.6 & 2533.6 & 3235.7 & & & \\
 			&  & \citet{lie_vibrational_1992} & 4088.9 & 2407.5 & 2533.4 & 3236.9 &  &  &  \\
		\hline \hline
		\end{tabular}
	\end{center}
	\tablenotetext{a}{This work unless otherwise specified.} 
	\tablenotetext{b}{See supplementary material of this reference.}
	\tablenotetext{c}{This value of \citet{Sanz-Sanz2013} is a discrepant by 19~cm$^{-1}$ with respect to our value, while all of their other frequencies agree to within less than 1~cm$^{-1}$ with ours.  This discrepancy is likley due to a misprint in their work, which also affects their corresponding ZPE value.}
	\tablenotetext{d}{This is the rotational ZPE value, taking into account the rotational excitation of the first allowed H$_3^+$ level which lies 64.12~cm$^{-1}$ above the vibrational ZPE of 4491.5~cm$^{-1}$.}
\end{table*}

Table~\ref{tab:theo1} gives the calculated energies for this energy profile, with the entrance energy subtracted out.  Our calculated BO electronic energy for Min$_1$ and Min$_2$ is $-1.8525663 E_{\rm h}$.  For TS$_3$, it is $-1.8383087 E_{\rm h}$.  Our ZPE-uncorrected results are from the calculated ic-MRCI(16 active orbitals)/aug-cc-pV5Z energy differences.  Our ZPE-corrected results add to this the ZPE energy differences, using the ZPE values given in Table~\ref{tab:theo2} and the rotational ZPE for H$_3^+$ (as explained below).  The barrier height, $E_{\rm b}$, for Reaction~(\ref{eq:D}) is highlighted in bold.  Also given in Table~\ref{tab:theo1} are the energies for both Min$_1$ and Min$_2$ from \citet{moyano_interpolated_2004} and \citet{Sanz-Sanz2013} and the exoergicity from \citet{ramanlal_deuterated_2004}.

The ZPE values have been derived using the vibrational frequencies $\upomega_s$ for each oscillating mode listed in Table~\ref{tab:theo2}.  The values of $\upomega_s$ have been calculated using the harmonic oscillator approximation.  Real values for $\upomega_s$ are listed in order of increasing frequency.  For TS$_3$, the corresponding imaginary frequency ($\upomega_{\rm im}$ discussed below) is given before the real $\upomega_s$ values.

The number of normal modes of an $N$-atom non-linear molecule is equal to $3N-6$.  Hence, there are 6 and 3 frequencies describing the vibrational motion of H$_4^+$ and H$_3^+$, respectively.  The energy for a vibrational level corresponding to the set of quantum numbers $v_1$, $v_2$,...,$v_{3N-6}$ is given, with respect to the bottom of the potential well, by the sum of the individual harmonic oscillator energies:

\begin{equation}\label{eq:theo1}
E\left(v_1,v_2,...,v_{3N-6}\right)= \sum_{s=1}^{3N-6} \hbar \upomega_s \left(v_s+\frac{1}{2}\right).  
\end{equation}

\noindent This expression provides the ZPE value when all of the oscillators $s$ are set to their $v_s=0$ ground state, giving

\begin{equation}\label{eq:theo2}
E_{\rm zp}= \frac{1}{2} \sum_{s=1}^{3N-6} \hbar \upomega_s.   
\end{equation}

\noindent In order to calculate the ZPE for transition states, the summation in Equation~(\ref{eq:theo2}) is limited to the real frequencies.  We also note that for H$_3^+$, we use the rotational ZPE, which takes into account the fact that the H$_3^+(J=K=0)$ ground rotational state is forbidden by the Pauli principle \citep{ramanlal_deuterated_2004}, where $J$ is the rotational angular momentum quantum number and $K$ is the projection of $J$ along the symmetry axis of the system.  The first allowed level, $J=K=1$, lies 64.12~cm$^{-1}$ above the nominal ground state \citep{morong_h3+_2009,pavanello_calibration-quality_2012,jaquet_investigation_2013}.  The rotational ZPE that we use here is the sum of the ground state ZPE and the energy of this first allowed level.

\begin{figure}[t!]
	\begin{center}
	\includegraphics[scale=0.33]{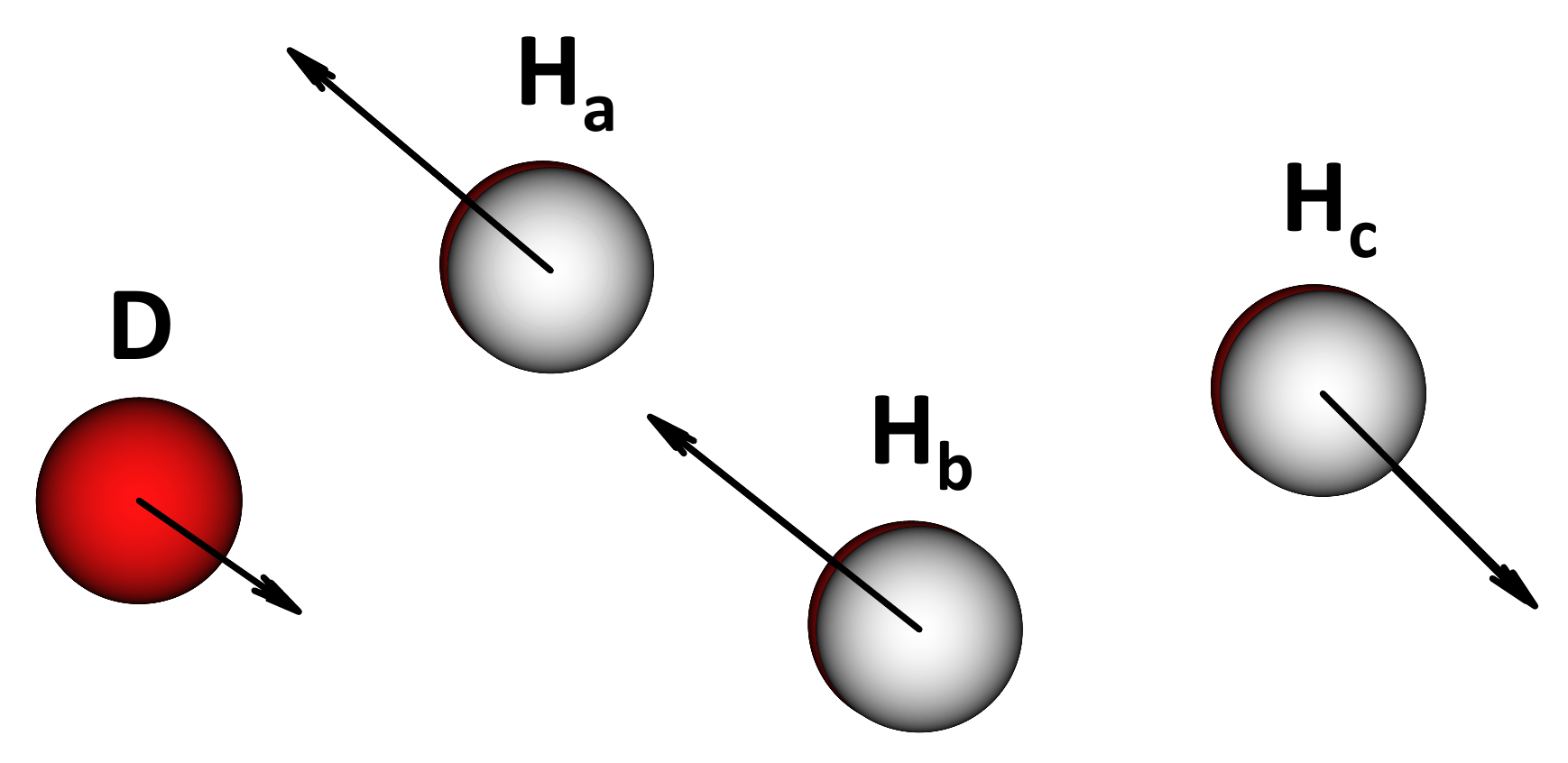}
	\caption{\label{fig:theo3} Atomic displacements of the mass-weighted normal mode coordinate of H$_3$D$^+$ corresponding to the imaginary frequency $\upomega_{\rm im}$ of TS$_3$.}
	\end{center}
\end{figure}

Particularly important for understanding Reaction~(\ref{eq:D}) is TS$_3$.  This is a first-order transition state and is thus characterized by a single imaginary frequency, $\upomega_{\rm im}$, highlighted in bold in Table~\ref{tab:theo2}.  The value of $\upomega_{\rm im}$ determines the negative curvature of the PES at the top of the reaction barrier. It is thus related to the barrier width and consequently to the tunneling probability, which we calculate in Section \ref{sec:tunneling}.  The normal coordinate associated with $\upomega_{\rm im}$ for H$_3$D$^+$ is depicted in Figure~\ref{fig:theo3}, illustrating the reaction coordinate TS$_3$ that connects Min$_1$ and Min$_2$.  Hydrogen atoms H$_{\rm a}$ and H$_{\rm b}$ are moving in parallel in the general direction towards the deuterium atom D, to form a triangular H$_2$D moiety.  Meanwhile, hydrogen atom H$_{\rm c}$ travels in the opposite direction to form the elongated H$_{\rm b}$-H$_{\rm c}$ bond of Min$_2$.  Inverting the direction of the arrows leads to Min$_1$ on the other side of the barrier, with the formation of the H$_{\rm a}$H$_{\rm b}$H$_{\rm c}$ triangle and the elongated D-H$_{\rm a}$ bond.

Comparing our H$_4^+$ ZPE-uncorrected energies to the previously published \textit{ab initio} calculations, we find excellent agreement with the work of \citet{Sanz-Sanz2013} using a similar level of theory, as shown in Table~\ref{tab:theo1}.  This largely confirms the convergence of our MRCI expansion.  We find poorer agreement with the results of \citet{moyano_interpolated_2004}.   There are no previously published results for H$_3$D$^+$ for us to compare to our ZPE-corrected energies.

For the vibrational frequencies, we again find excellent agreement with the H$_4^+$ results of \citet{Sanz-Sanz2013}, as can be seen in Table~\ref{tab:theo2}.  The agreement is to within $< 0.7$~cm$^{-1}$, apart from a probable misprint for one of their frequencies for Min$_{1,2}$ (see the footnote of Table~\ref{tab:theo2}).  Larger discrepancies are observed comparing to the calculations of \citet{Alijah2008}, which were carried out using a smaller number of active orbitals.  Even so, their results are only discrepant by $<2.7$~cm$^{-1}$ for all frequencies, except for $\upomega_{\rm im}$ for TS$_3$, which differs by 7.6~cm$^{-1}$.  Lastly, our frequencies for the ${\rm H}_3^+$ isotopologues agree to within $<2.5$~cm$^{-1}$ with the variational calculations of \citet{lie_vibrational_1992}.

\subsection{Further Theoretical Considerations}

\subsubsection{ZPE-Corrected Profiles}

The ZPE corrections to the characteristic energies of the collision profiles (see Table~\ref{tab:theo1}) are calculated with respect to the ZPE of the H$_3^+$ in the entrance channel. These corrections significantly alter the shape of the BO-energy profile, as shown in Figure~\ref{fig:theo2}. For X~=~H, the well depths and the barrier height are reduced by 21\% and 33\%, respectively.  For X~=~D, the depth of Min$_1$ and the barrier height are reduced by 16\% and 53\%, respectively, but the depth of Min$_2$ is almost unchanged. These changes arise from the ZPE values, which are ordered for X~=~H as 

\begin{equation}\label{eq:theo3}
E_{\rm zp} \left({\rm Min}_1 \right) = E_{\rm zp} \left({\rm Min}_2 \right) > E_{\rm zp}\left({\rm H}_3^+\right)  > E_{\rm zp}\left({\rm TS}_3\right),
\end{equation}

\noindent and for X~=~D as

\begin{equation}\label{eq:theo4}
E_{\rm zp} \left({\rm Min}_{1}\right) > E_{\rm zp}\left({\rm H}_3^+\right) \approx E_{\rm zp}\left(\rm{Min}_{2}\right) > E_{\rm zp}\left({\rm TS}_3\right). \\
\end{equation}

\noindent This ordering is a result of the decrease in the ZPE vs.\ molecular structure.  The ZPEs of the minima are high because of the strong 3-atom cycle, while those of the transition states are low, because of the weaker open structure.  ${\rm H}_3^+$ is an intermediate case, with one less hydrogen, but it also has a strong 3-atom cycle. Lastly, the case of Min$_2$ for X~=~D is fortuitous, a result of the deuteration effect, as explained below.

\subsubsection{Effect of Deuteration}

For X~=~H, the ZPE-corrected profile is symmetric with Min$_1$ and Min$_2$ being isoenergetic.  But for X~=~D, the profile is asymmetric, with a potential well deeper for Min$_2$ than for Min$_1$.  In general the energy of the vibrational motions of the H$_4^+$ isotopologues are decreased by deuteration.  This can be seen in the reduction of the ZPE for the minima of H$_3$D$^+$ compared to those for H$_4^+$.  The reduction is larger for Min$_2$, where the D atom affects the high-frequency vibrations of the 3-atom cycle, than it is for Min$_1$, where the D atom acts on the low-frequency vibrations of the weak \mbox{X-H} bond.  As a result, the ZPE value for Min$_2$ is fortuitously very close to the ZPE of ${\rm H}_3^+$.  This explains the quasi-equality of the BO and ZPE-corrected energies at the Min$_2$ position, as reported in Equation~(\ref{eq:theo4}) and shown in Figure~\ref{fig:theo2}.

Deuteration also generates the exoergicity, $\Delta E_{\rm zp}$, of the X~=~D reaction.  The resulting $\Delta E_{\rm zp}$, given in Table~\ref{tab:theo1}, is equal to the difference between the rotational ZPE for H$_3^+$ and the ZPE for H$_2$D$^+$.

\subsubsection{Anharmonic and Non-Adiabatic Effects}

Our calculated energy profiles provide insight into the collision dynamics of the ${\rm D + H_3^+}$ reaction system.  But to make the computations readily tractable, we have included neither the anharmonic effects in the ZPE calculations nor non-adiabatic corrections to the BO PESs.  Still, these approximations are expected to have only a small effect on the calculated stationary energies and exoergicity, leading to only insignificant changes in our understanding of the reaction dynamics.  The correction due to anharmonic effects amounts to 5\% for the value of the ZPE difference between the H$_3^+$ entrance channel and the H$_2$D$^+$ exit channel, as estimated by comparing our exoergicity to the anharmoic results of \citet[][see our Table \ref{tab:theo1}]{ramanlal_deuterated_2004}.  Non-adiabatic corrections to the BO PES are also estimated to be on the order of $\approx 10\%$.  These corrections introduce nuclear mass effects that are ignored within the BO approximation.  We expect that these would probably raise the barrier height slightly, as in the case of the ${\rm H + H_2}$ reaction, where an increase of about 7~meV is observed together with a narrowing of the barrier \citep{mielke_benchmark_2005}.

\begin{figure}[t]
	\begin{center}
		\includegraphics[width=1\columnwidth]{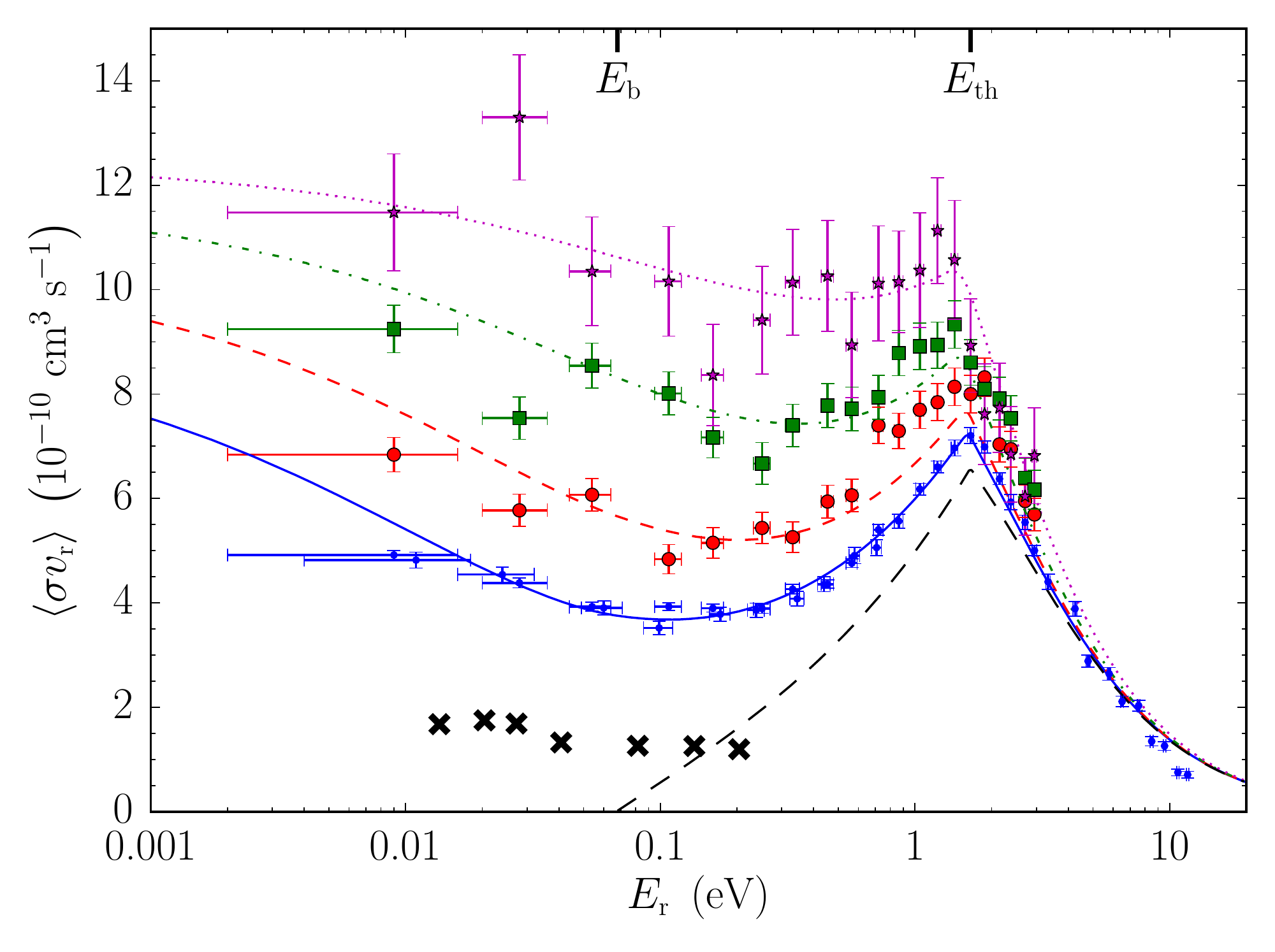} 
		\caption{Merged-beams rate coefficient, $\langle \sigma v_{\mathrm r} \rangle$, vs.\ the relative translational energy, $E_{\mathrm r}$, for a range of H$_3^+$-source pressures, $p_s$.  The vertical error bars represent the statistical uncertainty and the horizontal error bars show the energy spread at each $E_{\rm r}$.  The data correspond to $p_{\rm s} =$
			0.072~Torr (purple stars),	
			0.72~Torr (green squares),	
			0.36~Torr (red large circles), and
			0.48~Torr (blue small circles). 
		The lines show the results of the model given in Section \ref{sec:model} with a fitted H$_3^+$ internal temperature of 
			4400~K (purple dotted line), 
			2510~K (green dot-dashed line), 
			1610~K (red short-dashed line), and 
			1140~K (blue solid line). 
		The black long-dashed line is the inferred result for 0~K. The black crosses show theoretical data of \citet{moyano_interpolated_2004}. Vertical markers on the energy axis show the barrier height, $E_{\rm b}$, given in Table~\ref{tab:theo1}, and the threshold of the first competing channel, $E_{\rm th}$, given in Section~\ref{sec:chan}.}	
		\label{fig:mergedbeamsrate}
	\end{center}
\end{figure}

\section{Merged-Beams Rate Coefficient Results}
\label{sec:mergedbeamsrate}

Our measured $\langle \sigma v_{\rm r} \rangle$ vs.\ $E_{\rm r}$ is shown in Figure~\ref{fig:mergedbeamsrate}.  The results are given for
$E_{\rm r} \approx 0.01 - 10$~eV and for four different H$_3^+$-source pressures.  We attribute the decreasing trend from the highest $\langle \sigma v_{\rm r} \rangle$ data set to the lowest to be due to decreasing levels of H$_3^+$ internal excitation.  As the fraction of internally excited H$_3^+$ with energies sufficient to overcome $E_{\rm b}$ decreases, fewer ions can react and the measured $\langle \sigma v_{\rm r} \rangle$ correspondingly decreases.

We varied the H$_3^+$ internal excitation by adjusting the duoplasmatron operating parameters: source pressure, magnet current, filament current, and arc current.  Of these, $p_{\rm s}$ had the biggest influence on the measured $\langle \sigma v_{\rm r} \rangle$.  Variations in the magnet and filament currents around the typical operating conditions of 0.45~A and 12~A, respectively, had only a minor effect.  There was some influence of the arc current on $\langle \sigma v_{\rm r} \rangle$, but the variation was small between a setting of 1~A and our typical operating condition of 0.75~A.

\begin{figure}[t]
	\begin{center}
		\includegraphics[width=1\columnwidth]{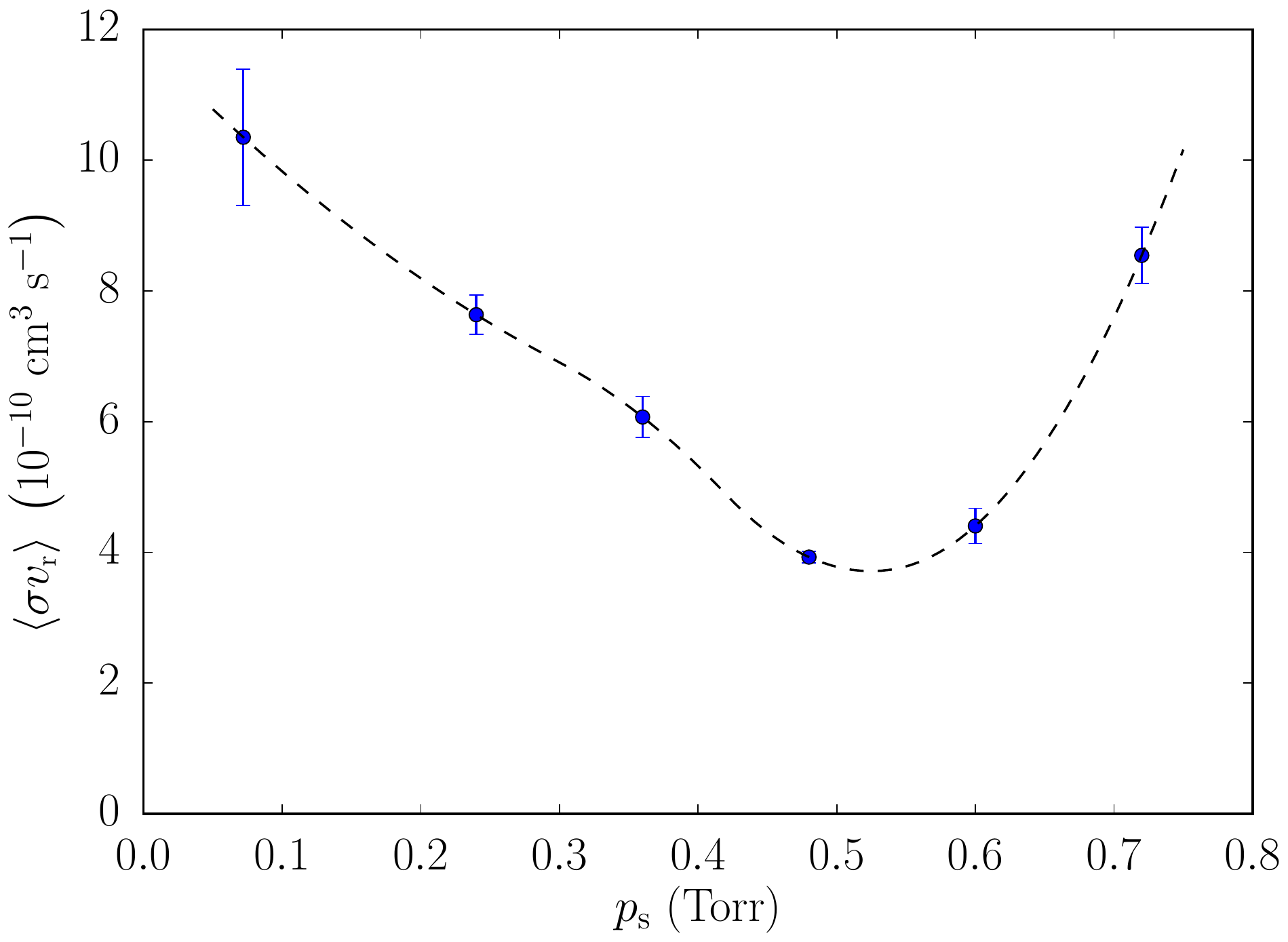} 
		\caption{Experimental merged-beams rate coefficient, $\langle \sigma v_{\mathrm r} \rangle$, for $E_{\rm r}=54\pm10$~meV vs.\ the pressure $p_{\rm s}$ inside the H$_3^+$-source. Shown are the measured data points with statistical errorbars. The dashed line is a quadratic interpolation of the data as guide to the eye.}
		\label{fig:press}
	\end{center}
\end{figure}

The observed dependence of the H$_3^+$ internal excitation on source pressure can be understood by considering the behavior of $\langle \sigma v_{\rm r} \rangle$ vs.\ $p_{\rm s}$, as shown in Figure~\ref{fig:press} for $E_{\rm r}= 54\pm10$~meV.  This collision energy is just below $E_{\rm b}$.  Classically, $\langle \sigma v_{\rm r} \rangle$ should be zero for cold H$_3^+$.  Taking tunneling into account, we expect $\langle \sigma v_{\rm r} \rangle \lesssim 4
\times 10^{-11}$~cm$^3$~s$^{-1}$ (see Section~\ref{sec:tunneling}).  However, we have measured $\langle \sigma v_{\rm r} \rangle \gg 4 \times 10^{-11}$~cm$^3$~s$^{-1}$.  Clearly the H$_3^+$ in our measurement is internally excited.  We attribute this to the gas-phase formation mechanism for the H$_3^+$, namely proton transfer between H$_2^+$ and H$_2$, with at least one of the two being vibrationally excited.  Past theoretical and experimental studies into the formation of H$_3^+$ have found internal energies of $\sim 0.5-1$~eV \citep[as reviewed in][]{oconnor_reaction_2015}.  

As for the pressure dependence shown in Figure~\ref{fig:press}, initially $\langle \sigma v_{\rm r} \rangle$ decreases with increasing $p_{\rm s}$.  We attribute this to collisions between H$_3^+$ and H$_2$ in the source that cool the H$_3^+$.  This collisional cooling increases with increasing source pressure.  Similar behavior has been seen in experimental photodissociation studies, which found generally decreasing levels of internal excitation with increasing source pressure (X.\ Urbain, private communication).  Then, at $p_{\rm s} \sim 0.5$~Torr, $\langle \sigma v_{\rm r} \rangle$ begins to increase with increasing source pressure.  We attribute this to collisional re-excitation of the accelerated H$_3^+$ as it passes through the residual gas downstream of the source.  As $p_{\rm s}$ increases, so does the H$_2$ streaming out of the duoplasmatron extraction aperture.  This increases the residual gas pressure downstream of the source, enabling collisional re-exciation to become important.  A similar mechanism has been proposed by \citet{kreckel_high-resolution_2010} to explain the heating of H$_3^+$ ions accelerated from a supersonic expansion ion source.  

Clearly, there is a minimal level of internal excitation of the H$_3^+$  which can be achieved with our current experimental configuration.  A setting of $p_{\rm s} \approx 0.48$~Torr appears to provide the lowest level of internal excitation for our H$_3^+$ ions.  The corresponding results are shown by the blue data points in Figure~\ref{fig:mergedbeamsrate} and listed in Table~\ref{tab:data1}.

\begin{table}[p]
	\begin{center}
		\caption{List of experimental merged-beams rate coefficients, $\langle
			\sigma v_{\mathrm r} \rangle$, with corresponding one-sigma
			statistical uncertainties, $\Delta \langle \sigma v_{\mathrm r}
			\rangle$, as a function of the relative translational energy, $E_{\rm
				r}$, with the one-sigma width of the collision-energy spread,
			$\Delta E_{\rm r}$, vs.\ applied floating cell voltages, $U_{\rm
				f}$. The data listed here corresponds to the measurement, where the internal temperature of the H$_3^+$ inferred from our model is 1140~K.}
		\label{tab:data1}
		\begin{tabular}{DDDDD}
			\hline
			\hline
			\multicolumn2c{$U_{\rm f}$} & 
			\multicolumn2c{$E_{\rm r}$} & 
			\multicolumn2c{$\Delta E_{\rm r}$} &
			\multicolumn2c{$\langle \sigma v_{\mathrm r} \rangle$} & 
			\multicolumn2c{$\Delta \langle \sigma v_{\mathrm r} \rangle$} \\
			\cmidrule(lr){3-6}\cmidrule(lr){7-10}
			\multicolumn2c{(kV)} & 
			\multicolumn4c{(eV)} &
			\multicolumn4c{$\left( 10^{-10}~{\rm cm}^3~{\rm s}^{-1}\right)$} \\
			\hline
			\decimals
			-0.900 &  10.774 & 0.120 & 0.754 & 0.066 \\
			-0.800 &  8.501  & 0.107 & 1.350 & 0.086 \\
			-0.700 &  6.506  & 0.093 & 2.112 & 0.102 \\
			-0.600 &  4.784  & 0.080 & 2.886 & 0.120 \\
			-0.500 &  3.333  & 0.066 & 4.404 & 0.145 \\
			-0.450 &  2.707  & 0.060 & 5.543 & 0.138 \\
			-0.400 &  2.148  & 0.053 & 6.377 & 0.111 \\
			-0.350 &  1.655  & 0.047 & 7.205 & 0.153 \\
			-0.300 &  1.227  & 0.040 & 6.604 & 0.113 \\
			-0.250 &  0.864  & 0.034 & 5.567 & 0.136 \\
			-0.225 &  0.707  & 0.031 & 5.059 & 0.154 \\
			-0.200 &  0.565  & 0.028 & 4.763 & 0.082 \\
			-0.175 &  0.440  & 0.025 & 4.358 & 0.143 \\
			-0.150 &  0.331  & 0.021 & 4.263 & 0.091 \\
			-0.125 &  0.238  & 0.018 & 3.861 & 0.136 \\
			-0.100 &  0.161  & 0.016 & 3.901 & 0.075 \\
			-0.075 &  0.099  & 0.013 & 3.521 & 0.129 \\
			-0.050 &  0.054  & 0.010 & 3.928 & 0.089 \\
			-0.025 &  0.024  & 0.008 & 4.543 & 0.146 \\
			0.000  &  0.009  & 0.007 & 4.918 & 0.083 \\
			0.025  &  0.011  & 0.007 & 4.820 & 0.152 \\
			0.050  &  0.028  & 0.008 & 4.383 & 0.094 \\
			0.075  &  0.060  & 0.011 & 3.905 & 0.137 \\
			0.100  &  0.108  & 0.013 & 3.930 & 0.075 \\
			0.125  &  0.172  & 0.016 & 3.782 & 0.136 \\
			0.150  &  0.251  & 0.019 & 3.891 & 0.089 \\
			0.175  &  0.345  & 0.022 & 4.079 & 0.139 \\
			0.200  &  0.454  & 0.025 & 4.359 & 0.079 \\
			0.225  &  0.579  & 0.028 & 4.908 & 0.156 \\
			0.250  &  0.719  & 0.031 & 5.402 & 0.104 \\
			0.300  &  1.045  & 0.037 & 6.179 & 0.111 \\
			0.350  &  1.430  & 0.043 & 6.968 & 0.153 \\
			0.400  &  1.876  & 0.049 & 6.988 & 0.118 \\
			0.450  &  2.381  & 0.055 & 5.932 & 0.145 \\
			0.500  &  2.946  & 0.061 & 5.003 & 0.102 \\
			0.600  &  4.251  & 0.073 & 3.889 & 0.140 \\
			0.700  &  5.788  & 0.084 & 2.642 & 0.122 \\
			0.800  &  7.556  & 0.096 & 2.033 & 0.103 \\
			0.900  &  9.550  & 0.108 & 1.262 & 0.834 \\
			1.000  &  11.769 & 0.119 & 0.710 & 0.064 \\
			\hline
		\end{tabular}
	\end{center}
\end{table}

\section{Discussion}
\label{sec:discussion}

\subsection{Competing Channels}
\label{sec:chan}

There are no exoergic channels to compete with Reaction~(\ref{eq:D}).  All of the competing channels are endoergic.  Up to the atomization limit, these include:

\begin{subequations}\label{eq:chan}
	\begin{align}
	{\rm D} + {\rm H}_3^+ & \rightarrow {\rm HD} + {\rm H}_2^+ - 1.65~{\rm eV}, \label{eq:chan:a}\\
	& \rightarrow {\rm HD}^+ + {\rm H}_2 - 1.67~{\rm eV}, \label{eq:chan:b}\\
	& \rightarrow {\rm HD} + {\rm H} + {\rm H}^+ - 4.30~{\rm eV}, \label{eq:chan:c}\\
	& \rightarrow {\rm H}_2 + {\rm D} + {\rm H}^+ - 4.34~{\rm eV}, \\
	& \rightarrow {\rm HD}^+ + {\rm H} + {\rm H} - 6.15~{\rm eV}, \\
	& \rightarrow {\rm H}_2^+ + {\rm D} + {\rm H} - 6.17~{\rm eV}, \\
	& \rightarrow {\rm D} + {\rm H} + {\rm H} + {\rm H}^+ - 8.82~{\rm eV}. \label{eq:chan:g}
	\end{align}
\end{subequations}

\noindent The threshold energies for these exoergic channels have been calculated using the dissociation energy of ${\rm H}_3^+$ from \citet{jaquet_investigation_2013}, the dissociation energies of diatomic molecules tabulated by \citet{huber_molecular_1979}, and the atomic electron binding energies of \citet{kramida_nist_2018}.

These competing channels explain one of the most dramatic features of our measured merged-beams rate coefficient, namely the rapid decrease starting near the threshold for the first two competing Channels~(\ref{eq:chan:a}) and (\ref{eq:chan:b}).  We define this energy as $E_{\rm th}=1.65$~eV.  A similar behavior at the opening up of competing channels was seen in our earlier measurements of ${\rm C + H_3^+}$ \citep{oconnor_reaction_2015} and ${\rm O + H_3^+}$ \citep{de_ruette_merged-beams_2016}.  As for the yet-higher-energy endoergic Channels~(\ref{eq:chan:c})-(\ref{eq:chan:g}), we see no clear change in the energy-dependent behavior of our results that would correspond to these channels opening up.

\subsection{Cross Section Model for the Experimental Results}
\label{sec:model}

We have developed a semi-empirical model to describe the experimental results shown in Figure~\ref{fig:mergedbeamsrate}.  This model accounts for the dominant features seen in our measurements: the inferred reaction barrier, the varying levels of H$_3^+$ internal excitation, and the opening up of competing exoergic channels.  We base our model, in part, on the Langevin-like formalism given by \citet{levine_molecular_2005} for a scattering event with a reaction barrier.  In this model, the cross section is given by $\sigma = \pi b^2$, where $b$ is the maximum impact parameter for which the reaction proceeds.  In addition, all reactions are assumed to occur with a probability of unity for all impact parameters equal to or smaller than $b$.

For the first part of the model, we assume that any internal excitation energy, $E_{\rm int}$, for a given level in H$_3^+$ is fully available to overcome any reaction barriers.  Thus, the reaction will go forward when the sum of $E_{\rm r}$ and $E_{\rm int}$ is sufficient to overcome the combined energies of the repulsive centrifugal barrier and the reaction barrier.  This gives

\begin{equation}
  E_{\rm r} + E_{\rm int} \ge \frac{E_{\rm r} b_{\rm b}^2}{R_{\rm b}^2} + E_{\rm b},
\end{equation}

\noindent where $b_{\rm b}$ is the impact factor taking the reaction barrier into account and $R_{\rm b}$ is the radial separation of the reactants at the location of the reaction barrier.  Solving for $b_{\rm b}^2$ gives

\begin{equation}
  b_{\rm b}^2 \le R^2_{\rm b}
  \left[1 + \frac{E_{\rm int} - E_{\rm b}}{E_{\rm r}}\right].
\end{equation}

\noindent We take the maximum value of $b_{\rm b}^2$.

For the second part of the model, we introduce a flux reduction factor, $S(E_{\rm r},E_{\rm int},E_{\rm th})$, to account for the opening of the competing exoergic channels discussed in Section~\ref{sec:chan}.  The value of $E_{\rm r}$ where the first competing channel opens up can be shifted from $E_{\rm th}$ towards lower energies by all or part of $E_{\rm int}$, depending on the fraction, $f$, of $E_{\rm int}$ that goes into overcoming the threshold for the competing exoergic channel.  By analogy with the so-called survival factor introduced for dissociative recombination studies by \citet{stroemholm_absolute_1993}, we can then write

\begin{eqnarray}\label{eq:flux}
& & S\left(E_{\rm r},E_{\rm int},E_{\rm th}\right) = \\
& &
\begin{dcases}
  1 & E_{\rm r} < E_{\rm th}-f E_{\rm int} \\
  \frac{1}{\left(1+a(E_{\rm r}-E_{\rm th}+f E_{\rm int})\right)^2} & E_{\rm r} \geq E_{\rm th}-f E_{\rm int},
\end{dcases}\nonumber
\end{eqnarray}

\noindent where $a$ and $f$ are adjustable parameters.  Putting together everything so far, we have
\begin{eqnarray}
  \label{eq:csbarrier}
  & & \sigma_{\rm b}(E_{\rm r},E_{\rm int}) = \\
  & &
  \begin{dcases}
    0 & E_{\rm r} + E_{\rm int} < E_{\rm b} \\
  \pi R_{\rm b}^2
  \left[1+\frac{E_{\rm int}-E_{\rm b}}{E_{\rm r}}\right] 
  S(E_{\rm r},E_{\rm int},E_{\rm th}) & E_{\rm r} + E_{\rm int} \ge E_{\rm b}.   
  \end{dcases}
  \nonumber
\end{eqnarray}

Next, we take into account that the upper limit for a reaction cross section is commonly assumed to be the classical Langevin value, $\sigma_{\rm L}$.  The Langevin cross section results from the combined effects of the attractive charge-induced dipole moment between the D and H$_3^+$ and the repulsive centrifugal barrier and is given by \citep{levine_molecular_2005}

\begin{equation}\label{eq:cslangevin}
\sigma_{\rm L}(E_{\rm r})= \pi e \left(\frac{2 \alpha_{\rm D}}{E_{\rm r}}\right)^{1/2}.
\end{equation}

\noindent Here $\alpha_{\rm D}$ is the static dipole polarizability of D.  This is given by \citet{schwerdtfeger_table_2018} as $\alpha_{\rm D} = 9 a_0^3 / (8\pi\epsilon_0)$, where $a_0$ is the Bohr radius and $\epsilon_0$ is the vacuum permittivity.

Solving Equations~(\ref{eq:csbarrier}) and (\ref{eq:cslangevin}), we find that for a given value of $E_{\rm int}$, $\sigma_{\rm b} > \sigma_{\rm L}$ for $E_{\rm r}$ below some energy that we define as $E_{\rm x}$.  As $E_{\rm int}$ increases, so does the value of $E_{\rm x}$.  To avoid these situations, we select the reaction cross section to be the smaller of $\sigma_{\rm b}$ and $\sigma_{\rm L}$.  In addition, we assume complete scrambling of the nuclei during the ${\rm D + H_3^+}$ reaction.  This is guided by the theoretical approach of \citet{hugo_h3+_2009} for isotopic variants of the ${\rm H_2 + H_3^+}$ reaction.  For the ${\rm D + H_3^+}$ reaction, only three of the asymptotic channels lead to the formation of H$_2$D$^+$.  The fourth outgoing channel leads to the formation of H$_3^+$.  To account for this, we introduce a factor of $3/4$ into our reaction cross section, giving 

\begin{equation}\label{eq:csmin}
  \sigma(E_{\rm r},E_{\rm int}) =
  \frac{3}{4}
  \min\left[\sigma_{\rm b}(E_{\rm r},E_{\rm int}),\sigma_{\rm L}(E_{\rm r})\right]. 
\end{equation}

Now, in order to compare this reaction cross section to our experimental results, we need to take into account the excitation energy of each H$_3^+$ level involved in the reaction.  We do this assuming that the H$_3^+$ levels follow a Boltzmann distribution,

\begin{equation}
g(E_{\rm int})=\exp(-E_{\rm int}/\langle E_{\rm int} \rangle),
\end{equation}

\noindent where $\langle E_{\rm int} \rangle$ is a function of the internal temperature $T_{\rm int}$ of the H$_3^+$ and is derived from the partition function $Z(T)$,

\begin{equation}
  \label{eq:averageE}
  \langle E_{\rm int} \rangle =
  k_{\rm B} T_{\rm int}^2 \frac{1}{Z} \frac{\partial Z}{\partial T_{\rm int}}.
\end{equation} 

\noindent We use the parameterization of $\langle E_{\rm int} \rangle$ vs.\ $T_{\rm int}$ given in \citet{kylanpaa_first-principles_2011}.

In the penultimate step of our model, we convolve Equation~(\ref{eq:csmin}) over $E_{\rm int}$.  The resulting model cross section is given by 

\begin{eqnarray}
  \label{eq:sigma_mod}
&\sigma_{\rm mod}&(E_{\rm r}, \langle E_{\rm int} \rangle)=\\
  & &\frac{1}{\langle E_{\rm int} \rangle}\int_0^\infty \sigma(E_{\rm r},E_{\rm int}) \exp(-E_{\rm int}/\langle E_{\rm int} \rangle)dE_{\rm int}.
  \nonumber
\end{eqnarray}

\noindent There are four adjustable parameters in our model cross section: $R_{\rm b}$, $\langle E_{\rm int} \rangle$, $a$, and $f$.  For the other values needed in Equation~(\ref{eq:sigma_mod}), we use $E_{\rm b}=67.98$~meV  from our {\it ab initio} calculations (see Section~\ref{sec:theoresults}) and $E_{\rm th}=1.65$~eV from the calculated energetics for the competing exoergic channels (see Section~\ref{sec:chan}).

Lastly, in order to compare to our measured merged-beams rate coefficient, we multiplied Equation~(\ref{eq:sigma_mod}) by $v_{\rm r}$ and varied the four adjustable parameters to best fit the experimental data.  Given the complexity of the model cross section and the lack of any clean analytic formula for the cross section, we carried out a by-eye fit, as opposed to a least-squares fit.  This is not expected to be an issue as our model is over-constrained by the data.  For $E_{\rm} \lesssim E_{\rm th}$, the magnitude and energy dependence of the data are determined by $R_{\rm b}$ and $\langle E_{\rm int} \rangle$.  Having fixed those two parameters, we then fit for $a$ and $f$ using the data for $E_{\rm r} \gtrsim E_{\rm th}$.  In each energy range, we fit for two free parameters using our four sets of measured data, thereby making the system over-constrained.  As an additional constraint, we required that the fits all use the same set of values for $R_{\rm b}$, $a$, and $f$ and only let $\langle E_{\rm int} \rangle$ vary between the fits to the four data sets.

Our semi-empirical model results are shown in Figure~\ref{fig:mergedbeamsrate}.  The model clearly demonstrates all of the major energy dependencies seen in the experimental data, namely: (i) a pronounced minimum in the merged-beams rate coefficient near $E_{\rm r} \sim 0.1$~eV, (ii) a distinct increase in the merged-beams rate coefficient from this energy until the opening of the competing exoergic reaction channels, (iii) a subsequent rapid decrease in merged-beams rate coefficient, and (iv) an overall increase of the merged beams rate coefficient with increasing $\langle E_{\rm int} \rangle$ of the H$_3^+$.

Commenting on the best-fit parameters, we found the best agreement between the measured data and our model for $R_{\rm b} = 2.53$~a$_0$.  This is relatively close to the geometry of TS$_3$, which has a distance of 2.87~a$_0$ between the D atom and center-of-mass of the H$_3^+$ moiety,
as deduced from the optimized geometry computed by \citet{Sanz-Sanz2013}.  The best-fit values of $\langle E_{\rm int} \rangle$ for the various source conditions are 0.185, 0.32, 0.65, and 1.5~eV, corresponding to $T_{\rm int} = 1140$, 1610, 2510, and 4400~K, respectively.  The case where we attribute an internal temperature of 4400~K to the reacting H$_3^+$ illustrates the uncertainty in our model, as this temperature is beyond the calculated 4000~K dissociation limit of H$_3^+$ \citep{kylanpaa_first-principles_2011}.  Nevertheless, the inferred range of H$_3^+$ temperatures is in a reasonable agreement with previous estimates from our measurements of C and O reacting with H$_3^+$ \citep{oconnor_reaction_2015,de_ruette_merged-beams_2016}.  Finally, the fall-off in the merged-beams rate coefficient that starts near $E_{\rm th}$ is best fit with $a=0.3$ and $f=0.2$.  The value for $f$ suggests that 20\% of $E_{\rm int}$ goes into overcoming the opening up of the competing exoergic channels, while 80\% is transferred into the daughter products.

\subsection{Comparison to Theoretical Cross Sections}
\label{sec:Comparison}

The classical trajectory (CT) cross section calculations of \citet{moyano_interpolated_2004} for Reaction~(\ref{eq:D}) are shown in Figure~\ref{fig:mergedbeamsrate}, multiplied by the values of $v_{\rm r}$ that correspond to their reported collision energies.  Surprisingly, the CT data are non-zero below $E_{\rm b}$.  It appears that Moyano et al.\ have only taken into account the ZPE of the initial H$_3^+$ and have not accounted for the important ZPE changes along the reaction path.  As a result, the reaction complex begins with sufficient energy to overcome the ZPE-uncorrected reaction barrier.  This leads to the observed unphysical prediction in the low-energy reaction dynamics, an issue known as the ``ZPE-leakage'' problem that is encountered in CT and quasi-classical trajectory (QCT) simulations \citep{Lu1989,Guo1996}.  This probably explains the unphysical CT results of Moyano et al., who predicted a nonzero cross section at energies below $E_{\rm b}$ for both X~=~H and D collisions.

Several different solutions for solving the ZPE leakage have been proposed \citep[][and references therein]{Guo1996,Lee2018}.  Among these solutions, the ring-polymer molecular dynamics (RPMD) approach \citep{Habershon2012} was recently applied with success to the D$^+$ + H$_2$ $\rightarrow$ HD + H$^+$ reaction \citep{Bhowmick2018}.  This could be an interesting alternative to CT or QCT calculations for Reaction~(\ref{eq:D}).

\subsection{Thermal and Translational Temperature Rate Coefficients}
\label{sec:thermal}

Using our semi-empirical model cross section, we can generate rate coefficients for thermal conditions where the translational temperature of the gas, $T_{\rm gas}$, and the internal temperature, $T_{\rm int}$, are equal.  However, the published theoretical rate coefficients are more appropriately compared to a translational temperature rate coefficient where $T_{\rm int}=0$~K.  Here, we present both rate coefficients.  We also present a theoretical correction to our thermal results to account for the effects of tunneling through the reaction barrier.

\subsubsection{Model Rate Coefficients}

Using our cross section model, the thermal rate coefficient is given by
\begin{eqnarray}
  \label{eq:thermal}
& & k_{\rm mod}(T)= \left(\frac{8}{\pi \mu k_{\rm B}^3 T^3}\right)^{1/2} \\
& & \times \int_0^\infty \sigma_{\rm mod}(E_{\rm r}, \langle E_{\rm int} \rangle) E_{\rm r} \exp\left(\frac{-E_{\rm r}}{k_{\rm B}T}\right)dE_{\rm r}, \nonumber
\end{eqnarray}

\noindent where $T = T_{\rm gas} = T_{\rm int}$.  The value of $\sigma_{\rm mod}$ is from Equation~(\ref{eq:sigma_mod}) using our best-fit values of $R_{\rm b} = 2.53$~a$_0$, $a=0.3$, and $f=0.2$.  $\langle E_{\rm int} \rangle$ is obtained from Equation~(\ref{eq:averageE}).

\begin{figure}[t!]
	\begin{center}
		\includegraphics[width=1\columnwidth]{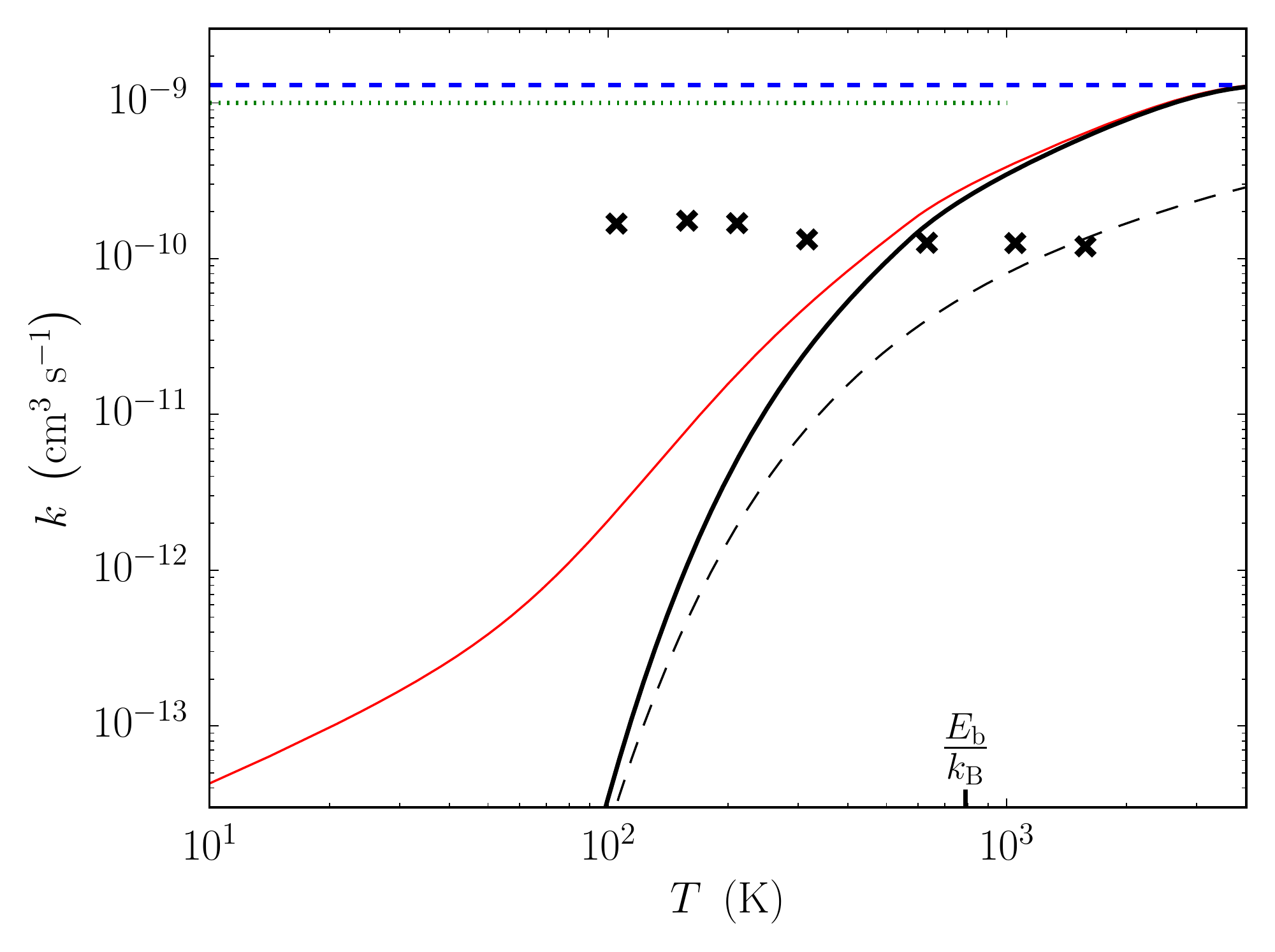} 
		\caption{Rate coefficient, $k$, vs.\ temperature, $T$.  Our model result from Equation~(\ref{eq:thermal}) is shown by the black thick line for the thermal case and by the black thin long-dashed line for the translational case (i.e., $T_{\rm int}=0$~K).  The tunneling-corrected thermal model from Equation~(\ref{eq:tunnel}) is shown by the red thin line.   The black crosses plot the CT calculations of \citet{moyano_interpolated_2004}.  The blue short-dashed line presents the Langevin rate coefficient from Equation~(\ref{eq:langevin}).  The green dotted line gives the rate coefficient commonly used in astrochemical models \citep{walmsley_complete_2004}.  Lastly, the energy of the reaction barrier, $E_{\rm b}/k_{\rm B}$, is indicated on the $T$ axis by the vertical marker.}
		\label{fig:thermal}
	\end{center}
\end{figure}

The resulting thermal rate coefficient is shown in Figure~\ref{fig:thermal} and given numerically in Table~\ref{tab:data2}.  The highest temperature presented corresponds to the thermal dissociation limit for H$_3^+$ \citep{kylanpaa_first-principles_2011}.  This upper limit corresponds to $E_{\rm r} = 0.34$~eV, which is well below $E_{\rm th}= 1.65$~eV for the competing exoergic channels.  Thus, the derived thermal rate coefficient is largely insensitive to the accuracy of the flux reduction factor $S\left(E_{\rm r},E_{\rm int},E_{\rm th}\right)$ of Equation~(\ref{eq:flux}).

As a self-consistency check, we also note that the thermal rate coefficient at
$T_{\rm int} = 1140$~K, corresponding to $\langle E_{\rm int} \rangle=0.185$~eV,
is nearly equal to the measured-merged beams rate coefficient for $E_{\rm r} = \langle E_{\rm int} \rangle = 0.185$~eV.  The thermal rate coefficient is $4.1\times 10^{-10}$~cm$^3$~s$^{-1}$ while the corresponding merged-beams rate coefficient is $\approx 3.7\times 10^{-10}$~cm$^3$~s$^{-1}$.

In order to enable ready implementation of our results into computational models, we have fit our model thermal rate coefficient with the commonly used Arrhenius-Kooij formula giving

\begin{eqnarray}
  \label{eq:arrfit}
  &&  k_{\rm mod}^{\rm fit}\left(T[{\rm K}]\right) = \\
  && 4.55\times 10^{-10}
  \left(\frac{T}{300}\right)^{0.5}
  \exp(-900/T)~{\rm cm}^3~{\rm s}^{-1}.
  \nonumber
\end{eqnarray}

\noindent This fit is accurate to within 5\% over the range $T = 100 - 4000$~K.  The lower limit is where the rate coefficient is $\approx 3 \times 10^{-14}$~cm$^3$~s$^{-1}$ and has been chosen as the rate coefficient rapidly decreases going to lower temperatures.  

We have also calculated the translational temperature rate coefficient, $k_{\rm tr}$, for the case where $T = T_{\rm gas}$ and $T_{\rm int} = 0$~K (i.e., $\langle E_{\rm int} \rangle = 0$~eV).  These data are plotted in Figure~\ref{fig:thermal} and listed in Table~\ref{tab:data2}.   

\begin{table}[p]
\begin{center}
\caption{Experimentally derived rate coefficients.}
\label{tab:data2}
\begin{tabular}{DDDD}
	\hline
	\hline
	\multicolumn2c{T}                    & \multicolumn2c{$k_{\rm tr}(T)$} & \multicolumn2c{$k_{\rm mod}(T)$} & \multicolumn2c{$k_{\rm tun}(T)$} \\
		\cline{3-8}
	\multicolumn2c{$\left({\rm K}\right)$} & \multicolumn6c{$\left( {\rm cm}^3~{\rm s}^{-1}\right)$\footnote{$a[-b]=a\times 10^{-b}$}}     \\
	\hline
	\decimals
	10.00	& ~       	   & ~       	& 4.274[-14] \\
	11.66	& ~       	   & ~       	& 5.093[-14] \\
	13.59	& ~       	   & ~       	& 6.091[-14] \\
	15.85	& ~       	   & ~       	& 7.316[-14] \\
	18.48	& ~       	   & ~       	& 8.834[-14] \\
	21.54	& ~       	   & ~       	& 1.073[-13] \\
	25.12	& ~       	   & ~       	& 1.314[-13] \\
	29.29	& ~       	   & ~       	& 1.624[-13] \\
	34.15	& ~       	   & ~       	& 2.033[-13] \\
	39.81	& ~       	   & ~       	& 2.589[-13] \\
	46.42	& ~       	   & ~       	& 3.374[-13] \\
	54.12	& ~       	   & ~       	& 4.528[-13] \\
	63.10	& ~       	   & ~       	& 6.289[-13] \\
	73.56	& ~       	   & ~       	& 9.061[-13] \\
	85.77	& ~       	   & ~       	& 1.352[-12] \\
	100.0	& 2.092[-14]   & 3.418[-14]	& 2.076[-12] \\
	116.6	& 6.944[-14]   & 1.269[-13]	& 3.251[-12] \\
	135.9	& 1.964[-13]   & 4.009[-13]	& 5.134[-12] \\
	158.5	& 4.845[-13]   & 1.096[-12]	& 8.085[-12] \\
	184.8	& 1.063[-12]   & 2.636[-12]	& 1.258[-11] \\
	215.4	& 2.107[-12]   & 5.664[-12]	& 1.919[-11] \\
	251.2	& 3.831[-12]   & 1.104[-11]	& 2.861[-11] \\
	292.9	& 6.468[-12]   & 1.979[-11]	& 4.165[-11] \\
	341.5	& 1.025[-11]   & 3.312[-11]	& 5.928[-11] \\
	398.1	& 1.537[-11]   & 5.243[-11]	& 8.273[-11] \\
	464.2	& 2.201[-11]   & 7.946[-11]	& 1.137[-10] \\
	541.2	& 3.027[-11]   & 1.166[-10]	& 1.545[-10] \\
	631.0	& 4.022[-11]   & 1.648[-10]	& 2.059[-10] \\
	735.6	& 5.189[-11]   & 2.187[-10]	& 2.609[-10] \\
	857.7	& 6.527[-11]   & 2.788[-10]	& 3.208[-10] \\
	1000	& 8.034[-11]   & 3.465[-10]	& 3.876[-10] \\
	1166	& 9.706[-11]   & 4.234[-10]	& 4.632[-10] \\
	1359	& 1.154[-10]   & 5.106[-10]	& 5.489[-10] \\
	1585	& 1.353[-10]   & 6.090[-10]	& 6.458[-10] \\
	1848	& 1.568[-10]   & 7.186[-10]	& 7.536[-10] \\
	2154	& 1.799[-10]   & 8.376[-10]	& 8.708[-10] \\
	2512	& 2.046[-10]   & 9.618[-10]	& 9.930[-10] \\
	2929	& 2.307[-10]   & 1.083[-9]	& 1.112[-9]  \\
	3415	& 2.582[-10]   & 1.190[-9]	& 1.216[-9]  \\
	3981	& 2.865[-10]   & 1.264[-9]	& 1.287[-9]  \\
	\hline
\end{tabular}
\end{center}
\end{table}

\subsubsection{Tunneling-Corrected Thermal Rate Coefficient}
\label{sec:tunneling}

In order to determine the influence of tunneling through the reaction barrier on the thermal rate coefficient for Reaction~(\ref{eq:D}), we use the analytic approximation of \citet{Eckart1930} as modified for an asymmetric barrier by \citet{Johnston:1966:Book}.  The tunneling-corrected thermal rate coefficient can then be written as

\begin{equation}
k_{\rm tun}(T) = \Gamma(T) k_{\rm mod}(T).
\end{equation}

\noindent Here $\Gamma(T)$ is the tunneling correction factor and can be expressed as

\begin{equation}
  \label{eq:tunnel}
  \Gamma(T) =
  \frac{1}{k_{\rm B}T}
  \int_{-V_{\rm f}} ^\infty  P(E_{\rm ts}) \exp\left(\frac{-E_{\rm ts}}{k_{\rm B} T}\right) dE_{\rm ts},
\end{equation}
where $P(E_{\rm ts})$ is the tunneling probability \citep{miller_tunneling_1979}.  The computed energy profile of the reaction and the corresponding normal mode frequencies of TS$_3$, 
described in Section~\ref{sec:theoresults}, provide a complete parameterization of the generalized Eckart potential, allowing us to express
$P(E_{\rm ts})$ in terms of the forward and reverse barrier heights ($V_{\rm f}$ and $V_{\rm r}$, respectively) and the magnitude of the imaginary frequency, $\upomega_{\rm b}=\left| \upomega_{\rm im} \right|$ (which quantifies the width of the reaction barrier of the transition state).  The quantity $E_{\rm ts} = E_{\rm r} + \langle E_{\rm int} \rangle -V_{\rm f}$ is the total reaction energy available to overcome the forward barrier of the transition state.  The value of $P(E_{\rm ts})$ can be evaluated as \citep{miller_tunneling_1979}

\begin{equation}
P(E_{\rm ts})=\frac{\sinh(A)\sinh(B)}{\sinh^2 \left(\frac{A+B}{2} \right)+\cosh^2 \left(C \right)},
\end{equation}

\noindent with $A$, $B$, and $C$ defined as

\begin{eqnarray}
A&=&\frac{4\pi}{\hbar \upomega_{\rm b}} \left(E_{\rm ts}+V_{\rm f}\right)^{1/2}\left(V_{\rm f}^{-1/2}+V_{\rm r}^{-1/2}\right)^{-1}\\
B&=&\frac{4\pi}{\hbar \upomega_{\rm b}} \left(E_{\rm ts}+V_{\rm r}\right)^{1/2}\left(V_{\rm f}^{-1/2}+V_{\rm r}^{-1/2}\right)^{-1}\\
C&=&2\pi\left(\frac{V_{\rm f} V_{\rm r}}{\hbar^2 \upomega_{\rm b}^2}-\frac{1}{16}\right)^{1/2}.
\end{eqnarray}

For evaluation of the tunneling-corrected thermal rate coefficient, we use our theoretical results given in Section~\ref{sec:theoresults}.  There we find $V_{\rm f} = E_{\rm b} = 67.98$~meV and $V_{\rm r} = E_{\rm b} + | \Delta E_{\rm zp}| = 125.78$~meV, based on the barrier height and the exoergicity of the exit channel given in Table~\ref{tab:theo1}.  The value of $\upomega_{\rm b} = 875.9~{\rm cm}^{-1}$ is given in Table~\ref{tab:theo2}.

Figure~\ref{fig:thermal} presents our tunneling-corrected thermal rate coefficient, which is also given numerically in Table~\ref{tab:data2}.  At the highest temperatures shown, the tunneling correction is unimportant and $k_{\rm tun}(T)$ converges to $k_{\rm mod}(T)$.  As is expected, the correction increases with decreasing temperature.  At $T = E_{\rm b}/k_{\rm B} = 789$~K, corresponding to the barrier energy, tunneling contributes $\approx 4 \times 10^{-11}$~cm$^3$~s$^{-1}$ or 17\% to the corrected thermal rate coefficient.  Based on the work of \citet{Schwartz1998}, we estimate that there is less than a factor of 2 uncertainty in the correction at this temperature.  Going to lower temperatures, at $T = 75$~K we find $k_{\rm tun}(T) \approx 10^{-12}$~cm$^3$~s$^{-1}$ and at 10~K, $k_{\rm tun} = 4.3\times10^{-14}~{\rm cm}^3~{\rm s}^{-1}$. Using the work of Schwartz et al.\ as a guide, we estimate that there is at least an order-of-magnitude uncertainty in our $k_{\rm tun}$ at these temperatures.  Schwartz et al.\ showed that the accuracy of the $\Gamma(T)$ factor can be increased by fitting the Eckart potential function to the PES of the transition state.  That level of theoretical complexity is beyond the scope of this paper.

Given the above caveats about the accuracy of the tunneling calculations, we have parameterized our results for $k_{\rm tun}(T)$ in units of cm$^3$~s$^{-1}$ as
\begin{eqnarray}
\label{eq:tunfit}
& &  k_{\rm tun}^{\rm fit}\left(T[{\rm K}]\right) = \\
& &
\begin{dcases}
3.3 \times 10^{-11} \left(\frac{T}{300}\right)^{2.73} \exp(28/T)   & 10 \leq T<180 \\
3.0 \times 10^{-10} \left(\frac{T}{300}\right)^{0.64} \exp(-560/T) & 180 \leq T \leq 4000.
\end{dcases}
\nonumber
\end{eqnarray}
The accuracy of the fit is better than 18\% over the given temperature ranges.

\subsubsection{Comparison to Theoretical Rate Coefficients}

In Figure~\ref{fig:thermal} we compare to various theoretical rate coefficients for Reaction~(\ref{eq:D}): the Langevin value, the value currently recommended by astrochemical modelers, and the CT result of \citet{moyano_interpolated_2004}.  All three of these are only translational temperature rate coefficients, as they do not take into account any possible internal excitation of the H$_3^+$.

The Langevin rate coefficient is calculated by integrating $\sigma_{\rm L} v_{\rm r}$ over a Maxwell-Boltzmann distribution, yielding

\begin{equation}
  \label{eq:kL}
  k_{\rm L} = 2 \pi e \left(\frac{\alpha_{\rm D}}{\mu}\right)^{1/2}.
\end{equation}

\noindent This value is temperature independent.  Taking into account that only three of the outgoing channels contribute to H$_2$D$^+$ formation, the Langevin rate coefficient for Reaction~(\ref{eq:D}) is

\begin{equation}
  \label{eq:langevin}
  k_{\rm L} = 1.3\times 10^{-9}~{\rm cm}^3~{\rm s}^{-1}. 
\end{equation}

\noindent The Langevin value clearly overestimates the rate coefficient for this reaction at all temperatures of astrochemical relevance.  Going to the high temperature limit shown in Figure~\ref{fig:thermal}, $k_{\rm mod}(T)$ converges to $k_{\rm L}$.  This is expected given our definition of the reaction cross section in Equation~(\ref{eq:csmin}).

The rate coefficient recommended for astrochemical modeling appears to have originated with the work of \citet{walmsley_complete_2004}.  Their value is

\begin{equation}
  \label{eq:walmsley}
  k_{\rm W} = 1.0\times 10^{-9}~{\rm cm}^3~{\rm s}^{-1},
\end{equation}

\noindent and is given for the temperature range of $10-1000$~K.  It is not clear how they derived this Langevin-like value, but their value clearly overestimates the rate coefficient at astrochemically relevant temperatures.

Lastly, we have used the CT results of \citet{moyano_interpolated_2004} for ground state H$_3^+$ to generate a translational temperature rate coefficient.  We do this by multiplying their cross section data, calculated for $T_{\rm int}=0$~K, by $v_{\rm r}$ and plotting their monoenergetic results at the temperatures given by $T = \frac{2E_{\rm r}}{3k_{\rm B}}$.  The results are nearly an order of magnitude below $k_{\rm L}$ and approximately constant with temperature.  Compared to $k_{\rm mod}$, the Moyano et al.\ results overestimate the rate coefficient at low temperatures.  This is most likely due to the ZPE-leakage issue discussed in Section~\ref{sec:Comparison}.  At higher temperatures ZPE-leakage should cease to be an issue.  At these temperatures, their results are, not surprisingly, significantly below $k_{\rm mod}$, but they are in rough agreement with our results for $k_{\rm tr}$.  

\subsection{Astrophysical Implications} 

Our combined experimental and theoretical results indicate that Reaction~(\ref{eq:D}) proceeds at prestellar core temperatures of $\sim 10-20$~K with a rate coefficient of $ \lesssim 10^{-13}$~cm$^3$~s$^{-1}$.  This low rate coefficient arises from tunneling through a reaction barrier of $\approx 68$~meV.  Given the height of this barrier, we expect that the magnitude of the rate coefficient will be insensitive to the ortho-to-para ratio of H$_3^+$.  The lowest energy level of ortho-H$_3^+$ lies only 2.8~meV above the lowest allowed level for para-H$_3^+$ \citep{hugo_h3+_2009}.  This is an insignificant difference with respect to the reaction barrier energy.

Deuterated astrochemical models currently assume a rate coefficient for Reaction~(\ref{eq:D}) of $\sim 1 \times 10^{-9}$~cm$^3$~s$^{-1}$
\citep{roberts_modelling_2000,walmsley_complete_2004,albertsson_new_2013,majumdar_chemistry_2017}.  These should be updated to use our results presented here, but we expect that the result will be to essentially turn off this channel for deuterating H$_3^+$ at prestellar core temperatures.  Thus, current astrochemical models are likely to overestimate the H$_2$D$^+$ number density, $n({\rm H_2D^+})$.  In addition, this implies that HD is the primary species responsible for deuterating H$_3^+$ at these low temperatures \citep{hugo_h3+_2009,albertsson_new_2013,sipila_species--species_2017}
via

\begin{equation}
  \label{eq:HD}
  {\rm HD + H_3^+ \to H_2D^+ + H_2}.
\end{equation}

\noindent This reaction is barrierless and exoergic by 19.98~meV for the reactants and products in their lowest energy states \citep{ramanlal_deuterated_2004,hugo_h3+_2009}.  We are unaware of any publicly available deuterated astrochemical models and so our discussions here are purely qualitative.

Our current understanding of collapsing low- and high-mass prestellar cores may also be affected by Reaction~(\ref{eq:D}) being essentially closed below 20~K.  Ground-based observations of the deuterium fractionation ratio $D_{\rm frac}^{\rm N_2H^+} \equiv n({\rm N_2D^+}) / n({\rm N_2H^+)}$ are used as a chemical clock for comparing to dynamical models of core formation and evolution \citep{kong_deuterium_2015}.  These two ions are predicted to form primarily from the reactions

\begin{eqnarray}
  {\rm H_2D^+ + N_2} & \to & {\rm H_2 + N_2D^+}, \\
  {\rm H_3^+ + N_2}  & \to & {\rm H_2 + N_2H^+}.
\end{eqnarray}

\noindent Deuterium fractionation increases with time as a core begins to collapse.  It then decreases once a protostar forms and begins to heat the gas, enabling the endoergic reverse of Reaction~(\ref{eq:HD}) to take place, thereby reducing the formation of N$_2$D$^+$.

\citet{kong_deuterium_2015} report that values of $D_{\rm frac}^{\rm N_2H^+} \gtrsim 0.1$ are commonly observed in low- and high-mass prestellar cores.  These values imply chemical timescales longer than the local free-fall timescale given by simple gravitational collapse models.  Kong et al.\ posit that this is indirect evidence that magnetic fields play an important role in regulating the evolution and collapse of prestellar cores, and thereby of star formation.  Given that Reaction~(\ref{eq:D}) is essentially closed, reducing the H$_2$D$^+$ abundance, this implies that even longer chemical timescales are required to match the observed values of $D_{\rm frac}^{\rm N_2H^+}$.  If this is the case, then that strengthens the argument of Kong et al.

\section{Summary}
\label{sec:summary}

We have reported here a combined experimental and theoretical study of atomic D reacting with H$_3^+$, leading to the formation of H$_2$D$^+$.  Our findings indicate that this reaction is essentially closed at the $\sim 10-20$~K temperatures of astrochemical relevance.  We have presented thermal rate coefficients so that deuterated astrochemical models can be readily updated accordingly.

\acknowledgments

The authors thank Evelyne~Roueff and Fabrice Dayou for stimulating conversations. This research was supported, in part, by the NSF Division of Astronomical Sciences Astronomy \& Astrophysics Grants program under AST-1613267. P.-M.~H.\ was supported, in part, by the Deutsche Forschungsgemeinschaft (DFG) under grant number HI~2009/1-1. J.~L.\ thanks the ULB/VUB computing center and the Consortium des Equipements de Calcul Intensif (FRS-FNRS and Walloon Region) for computational support.  X.~U.\ is a Senior Research Associate of the Fonds de la Recherche Scientifique-FNRS and acknowledges travel support from Fonds de la Recherche Scientifique-FNRS through IISN grant number 4.4504.10.

\software{Molpro \citep{Werner2012,MOLPRO}}

\bibliography{D+H3+}

\end{document}